\documentstyle[aps,preprint,tighten,floats,epsf,rotate]{revtex}

\begin{document}
\draft
%%%%%%%%%%%%%%%%%%%%%%%%%%%%%%%%%%%%%%%%%%%%%%%%%%%%%%%%%%%%%%%%%%%%%%%%%

%\preprint{\vbox{\it 
%                        \null\hfill\rm    IP-BBSR/2001-2, Jan.'2001}\\\\}
%%%%%%%%%%%%%%%%%%%%%%%%%%%%%%%%%%%%%%%%%%%%%%%%%%%%%%%%%%%%%%%%%%%%%%%%%
%
\title{Formation and Collapse of False Vacuum Bubbles in 
Relativistic Heavy-Ion Collisions}
\author{Rajarshi Ray \footnote{e-mail: rajarshi@iopb.res.in},
Soma Sanyal \footnote{e-mail: sanyal@iopb.res.in}, and Ajit M.
Srivastava \footnote{email: ajit@iopb.res.in}}
\address{Institute of Physics, Sachivalaya Marg, Bhubaneswar 751005, 
India}
%
%\date{June 2002}
%
\maketitle
\widetext
\parshape=1 0.75in 5.5in
\begin{abstract}

 It is possible that under certain situations, in a relativistic heavy-ion 
collision, partons may expand out forming a shell like structure. We analyze 
the process of hadronization in such a picture for the case when the
quark-hadron transition is of first order, and argue that the inside region 
of such a shell must correspond to a supercooled (to $T = 0$) deconfined
vacuum. Hadrons from that region escape out, leaving a bubble of pure 
deconfined vacuum with large vacuum energy. This bubble undergoes 
relativistic collapse, with highly Lorentz contracted bubble walls, and may
concentrate the entire energy into extremely small regions. Eventually 
different portions of bubble 
wall collide, with the energy being released in the form of particle 
production. Thermalization of this system can lead to very high
temperatures. With a reasonably conservative set of parameters, at
LHC, the temperature of the hot spot can reach as high as 3 GeV, and
well above it with more optimistic parameters. Such a hot spot can
leave signals like large $P_T$ partons, dileptons, and enhanced 
production of heavy quarks. We also briefly discuss a speculative 
possibility where the electroweak symmetry may get restored in the
highly dense region resulting from the decay of the bubble wall
via the phenomenon of non-thermal symmetry restoration (which is
usually employed in models of pre-heating after inflation). If that 
could happen then the possibility may arise of observing sphaleron 
induced baryon number violation in relativistic heavy-ion collisions.

\end{abstract}
\vskip 0.125 in
\parshape=1 0.75in 5.5in
%Key words: {quark-hadron transition, baryon violation, relativistic
%heavy-ion collisions}
%\newpage
%\begin{multicols}{2}
\narrowtext
%%%%%%%%%%%%%%%%%%%

\section{Introduction}

 There are strong reasons to  believe that in ultra-relativistic collisions 
of heavy nuclei, a hot dense region of quark-gluon plasma (QGP) may get 
created. There is a wealth of data which strongly suggests that already at 
CERN SPS this transient QGP state may have been achieved \cite{qgp}. With 
new data already coming out from RHIC at BNL \cite{rhic}, it may be only a 
matter of time that conclusive evidence of QGP detection would be obtained. 
Certainly, one expects that at LHC (in next few years) QGP will be produced
routinely. There is no question that detection of this new phase
of matter will be of utmost importance, not just for testing predictions
of QCD, but also in providing us glimpses of how our universe may have
looked like at few microseconds of age. It seems to be an appropriate 
stage that one starts looking beyond the detection of the QGP phase.
Many interesting possibilities have been discussed in the literature about
physical processes which may become observable once QGP is produced
in laboratory. For example, one will have the opportunity of studying
phase transitions in a relativistic field theory system under controlled 
laboratory situations. The richness of QCD phase diagram may become
available for probing with quark-hadron transitions occurring in these
experiments \cite{crtcl}. 

 In this paper we propose a novel possibility of forming a pure false
vacuum bubble in a relativistic heavy-ion collision, whose collapse 
and decay can lead to extremely hot, tiny regions, with temperatures
well above the initial temperature of the QGP system. We assume that 
the confinement-deconfinement phase transition is of first order, and
show that such a false vacuum bubble will 
form if the parton system (formed in the
collision) expands out in a shell like structure, leaving behind the 
supercooled deconfined phase with large vacuum energy density. Eventually, 
as the partons hadronize and escape out, the interface, separating this 
false vacuum bubble from the confined phase outside, undergoes relativistic 
collapse. Due to absence of any plasma inside, the motion of this interface 
becomes ultra-relativistic, with highly Lorentz contracted interface 
thickness. Entire energy of the false vacuum bubble gets converted to the 
kinetic energy of the wall.  Eventually different portions of bubble wall 
collide, converting the entire energy into particles. These particles may 
thermalize and lead to an extremely hot, very tiny region (apart from
possible effects of quantum fluctuations of the bubble wall, as we
discuss later). We find that even at RHIC energies, the temperature
at these hot spots can be more than 1 GeV. For LHC the temperature can reach
several  GeVs.  This will have important signals such as increased 
production of heavy quarks, anomalously large values of $P_T$ of some 
hadrons (those coming from the hot spot) etc. Also, the net energy of the 
false vacuum bubble may be very large even at RHIC ($\sim$ 1 TeV),
hence the possibility of Higgs and top quark production may also arise
(depending on the decay products of the bubble wall). 

 At LHC energies, the net energy of this false vacuum bubble 
is very large which should lead to very dense parton system after 
the bubble wall decay. Due to initial
very large density of partons at such a spot (forming a
non-equilibrium system) we speculate on the possibility that
electroweak symmetry may get restored there via the phenomenon of
non-thermal symmetry restoration, as in the models of pre-heating
after inflation \cite{rsnc}. If that could happen then it
raises the possibility that sphaleron transitions 
may occur in these regions (depending on the size of such a region), which 
will lead to baryon number non-conservation.  It is needless to say that 
any possibility of detection of electroweak baryon number violation in 
laboratory experiments deserves attention, especially with its 
implications for the theories of baryogenesis in the early universe. 
We mention here that the possibility of baryon number violation 
in collider experiments has been discussed earlier, see, ref.\cite{ewstr}
and references therein. However, the discussion in these works
was about possibility of baryon number violation at high energies,
and not due to electroweak symmetry restoration. It is fair to say
that at this stage it is not clear whether it is possible to get
baryon number violating interactions at high energies. However, baryon 
number violation in the electroweak symmetric phase is on 
rather strong foundations \cite{sphl0,sphl1,sphl2}, and our reference
is to this type of baryon number violation. 

 The paper is organized as follows. In section II, the basic physical
picture of the model is discussed where it is argued how a shell
enclosing deconfined vacuum may form in ultra-relativistic heavy-ion
collisions. Section III discusses the properties of the shell, the 
surface energy, volume energy etc. Section IV discusses the evolution
of this shell and its final ultra-relativistic collapse leading to
high concentration of energy in a tiny region. In section V we discuss
the issue of thermalization of the decay products of bubble wall. 
Section VI presents
results for various ranges of parameters for LHC and RHIC energies,
as well as for the possibility of non-thermal restoration of the
electroweak symmetry. Discussion of results and the conclusion is 
presented in section VII.

\section{Physical Picture of the Model}

 The main aspect of our model is based on the observation \cite{shel} 
that under certain situations, the expanding parton 
system may form a shell like structure. This type of picture emerges under a
variety of conditions. For example, in a hydrodynamical expansion, the 
rarefaction wave reaching center gives rise to a shell like structure 
\cite{shel}. (We mention, that shell like structure with a {\it maximum}
of density at center has also been discussed in the literature \cite{nut}.)
It has been shown that shell structure will arise generically in an expanding 
parton system when partons are ultra-relativistic and particle collisions are 
not dominant \cite{shel}. This happens for the simple reason that with 
all partons having velocity $\simeq c$, the partons pile up in a shell of 
radius $\simeq c\times t$, with the thickness of the shell being of the order 
of the size of the initial region. Indeed, the original so called {\it baked
Alaska} model for the disoriented chiral condensates, proposed by
Bjorken et al. \cite{dcc} utilizes a shell like picture for the expanding 
parton systems (see also, ref.\cite{bblch} where a similar picture has been
discussed). Our model is based on this generic shell structure 
of the parton system. 

  However, there is one important difference between our model and other 
discussions in the literature where a shell like structure of partons has 
been discussed. In refs.\cite{dcc,bblch}, the discussion was in the context 
of a second order chiral phase transition where the larger potential energy 
of the inside region slowly decreases as the chiral field relaxes towards 
the true vacuum. In contrast, our model is based on a first order 
confinement-deconfinement phase transition, where the vacuum energy 
of the inside region can only decrease by collapse of the interface, or 
via nucleation of true vacuum bubbles in this region, which we will argue 
to be very suppressed. Though we have presented the discussion in 
terms of a first order deconfinement-confinement transition, our entire 
discussion will also be valid for a first order chiral phase transition. 
Basic energy scales of these two transitions being roughly of same order, 
even quantitative aspects of the discussion will not change much. In 
discussions of \cite{shel} it is argued that the hadronization
in the shell will proceed from the inner boundary of the shell, as well
as the outer boundary of the shell. This assumes that the {\it empty}
region inside the shell will be in the confined vacuum. We argue below
that there is no reason to expect that. Rather, one expects that the shell
of partons will enclose a region of deconfined vacuum.

 Let us start with the initial stage of central collision of two nuclei. For 
very early stages after the collision, the region between the two receding
nuclei will be populated by a dense system of partons, with a temperature
which is expected to be well above the critical temperature $T_c$ of
deconfinement-confinement phase transition (at RHIC and LHC). (We will 
consistently refer to the phase transition as the deconfinement-confinement 
(D-C) phase transition rather than using the conventional terminology of 
quark-hadron transition. This is 
because we will utilize the difference between the vacua of the two phases of 
QCD, irrespective of the fact whether these vacua are populated by 
quark-gluon, or hadron degrees of freedom.) Subsequently, plasma will expand 
longitudinally for early times (for $\tau < R_A$, where $\tau$ is the proper 
time and $R_A$ is the radius of the nucleus), and will undergo 
three-dimensional expansion for larger times. For heavy nuclei with large $A$, 
and for center of mass energies at RHIC and LHC, it is expected that the 
plasma will undergo deconfinement-confinement phase transition during this
three dimensional expansion stage, and later the expanding hadronic
system will freezeout. In various models of freezeout, it is argued that
the freezeout happens soon after the D-C phase transition and that this stage 
is achieved at proper time hypersurface which is close to the hypersurface
when three-dimensional expansion commences \cite{frz1} (see, also ref. 
\cite{frz2}).

  The expected value of central energy density $\epsilon_i$ at the initial 
stage (hence the initial temperature $T_i$), increases with $A$ and with
$\sqrt{s}$. An estimate of the dependence of $\epsilon_i$ on $A$ and $\sqrt{s}$
can be obtained from the scaling relations proposed in ref.\cite{scl1} (see
also, ref. \cite{scl2}),

\begin{equation}
\epsilon_i  = 0.103 A^{0.504} (\sqrt{s})^{0.786} ~ {\rm GeV fm}^{-3},
\end{equation}

\noindent where $\sqrt{s}$ is in GeV. These scaling exponents reproduce the 
expected values of initial energy density for SPS, RHIC and LHC. Thus, 
for LHC, with $A$ = 208 and $\sqrt{s} = 5.5$ TeV (we will use $\sqrt{s}$ 
to refer to $\sqrt{s}$ per nucleon), we find $\epsilon_i \simeq 1.3 
 ~ {\rm TeV/fm}^3$ which is within the range of other estimates \cite{lhc}.

 We note that dependence on $A$ is somewhat weaker than the dependence
on $\sqrt{s}$. Thus, if one decreases value of $A$ and increases $\sqrt{s}$
suitably, one can still get large $\epsilon_i$ (suitable for having a QGP 
state initially at temperatures well above $T_c$). However, with lower value 
of $A$, $R_A$ will be smaller ($R_A \simeq 1.1 A^{1/3}$), implying that the
three-dimensional expansion will commence at earlier values of proper time.
This will have two effects. Temperature will decrease faster ($T \sim 
\tau^{-1}$ for three-dimensional expansion while $T \sim \tau^{-1/3}$ for
longitudinal expansion \cite{bj}). More importantly, with rapidly expanding
system, thermodynamic equilibrium will be maintained for a shorter duration 
of proper time, implying that freezeout will happen earlier. In principle,
there is no reason why freezeout cannot precede the D-C phase transition. 
In such a situation, initial thermalized QGP system expands longitudinally
for a short time, then undergoes three dimensional expansion. Due to rapid
3-dimensional expansion, the system falls out of equilibrium and keeps 
expanding as a parton system in a non-equilibrium 
state (still with high density 
of partons so that no hadronization takes place yet). Eventually the parton 
system will be dilute enough to enter into the non-perturbative regime,
and will undergo hadronization. This process of  hadronization will be
in a non-equilibrium state, where a parton system out of equilibrium converts 
to a hadron system (again, out of equilibrium). Possibility of 
freezeout preceding hadronization has also been discussed previously in
ref.\cite{frzph}.  A quantitative discussion of 
the limiting values of $\sqrt{s}$, and that of $A$ below which
freezeout may precede hadronization, can only be given using 
elaborate numerical computations (such as those in ref.\cite{frz1}).

  In such a picture of expansion of parton system, it is possible to
have the central region (in the shell) depleted of partons while the
partons get accumulated in the shell. Thus, central region {\it does not
cool} to become zero temperature, zero density QCD matter, as would
happen if the system was always in thermal equilibrium (and that would
then require that the interior of the shell be in the confined vacuum,
as assumed in \cite{shel}). Rather, in the above picture, the
interior of the shell is depleted of partons because of rapid expansion of 
the parton system. As thermal equilibrium is not maintained, parton collision 
rate lags behind the expansion rate. As mentioned above, in such a situation 
the analysis of refs. \cite{shel} suggests that partons, due to their 
relativistic velocities, pile up in a shell of radius $R \simeq c \times t$ 
where $t$  is the time in the center of mass frame of the collision. The 
thickness of the shell $\Delta r$ will be expected to be of the order of the 
size of the region when thermal equilibrium is first broken such that parton
system undergoes almost free expansion after that. 

  We mention here that it is conceivable that the initial parton system never 
thermalizes (for example for very small $A$), but nevertheless, deconfinement 
phase is achieved.  This possibility can originate from the discussions of 
non-thermal symmetry restoration in the context of inflation in the early 
universe \cite{rsnc}.  In these works it is shown that rapid particle 
production due to parametric resonance can lead to modification of the 
effective potential, even without any notion of thermal equilibrium. Under 
certain situations the symmetry restoration can be achieved even when thermal 
symmetry restoration is not expected. The early stages in nucleus-nucleus 
collisions resemble the stage of parametric resonance at least in the sense 
of rapid particle production. It is then possible that in this case also
deconfinement phase is achieved with parton system always remaining
out of equilibrium. This system will then undergo expansion forming a 
shell structure \cite{shel} before hadronizing eventually. (Later,
in Sec.VI, we will discuss whether a similar picture can also
emerge for the decay products of bubble wall, leading to non-thermal
restoration of the electroweak symmetry even when the system remains
in a non-equilibrium state.)

 Thus, we assume that the expanding parton system forms a shell, with
the interior of the shell being in the deconfined vacuum. At the outer
boundary of the shell there must be an interface separating the 
deconfining vacuum in the interior from the confining vacuum in the 
outer region \cite{dgl}. (We are assuming the D-C transition to be of first 
order, with a barrier separating the metastable deconfined vacuum from
the true confining vacuum, even at temperatures approaching zero. We will 
discuss this issue in detail below.) Partons will be piled up inside the 
shell, within a thickness $\Delta r$ of the interface. The center of the 
shell interior thus represents a supercooled (to $T = 0$) region trapped 
in the metastable deconfining vacuum. 

  It is somewhat uncommon to talk about zero temperature deconfining
phase of QCD. Normally one associates deconfinement phase with the 
quark-gluon plasma at temperatures above $T_c$. However, it is important
to distinguish the vacuum of a theory from the particle excitations
about that vacuum. Confining phase and the deconfining phase of QCD
should be taken to correspond to the two different vacua of some
effective theory of QCD, with the deconfining
vacuum being metastable at zero temperature. The two vacua may be 
characterized by an order parameter, such as the expectation value of the 
Polyakov line \cite{poly}. That means that certain 
type of gauge field background leads to one vacuum, while a different gauge 
field background gives the different vacuum. These two vacua are thus defined 
irrespective of the presence or absence of quarks and gluons, or
hadrons, as particle degrees of freedom (except as test particles 
in order to define confinement). It then makes sense to talk about 
deconfining vacuum without any quark or gluon being present as
particle degrees of freedom, though this vacuum will be metastable. (For 
example, if the chiral phase transition is of first order, with a 
chirally symmetric vacuum even at $T = 0$, then it will make perfect
sense to talk about a supercooled chirally symmetric phase with $T = 0$.)
 
 The type of picture we are taking in our model, with the shell interior
depleted of quarks and gluons but still trapped in the metastable deconfining 
vacuum, is very similar to what happens in the inflationary theories of the 
universe \cite{infl}. For example, in the old inflation, the scalar
field $\phi$ gets trapped in the false vacuum and supercools as the universe
undergoes exponential expansion due to non-zero potential energy of
$\phi$. Initially the universe is filled with scalar particles (as well as 
other particles) in thermal equilibrium corresponding to a temperature above 
the transition temperature. However, due to exponential expansion, initially 
the system cools maintaining equilibrium, but then falls out of equilibrium 
with all particles undergoing depletion due to exponential expansion of the 
universe. Eventually the particles are completely diluted away, while
the universe still remains in the metastable vacuum. This is precisely
the same picture as we are using in our model. The only thing required
for such a picture is that QCD admit a metastable vacuum even at zero 
temperature. The exponential expansion of the universe is replaced by 
rapid outward expansion of partons in our case. Possibility of supercooling 
and a brief inflationary phase at quark-hadron transition in the universe 
has also  been discussed in the literature \cite{qcdinfl}.

\section{Physical Properties of the Shell}

It is perfectly sensible to believe that QCD admits a metastable
vacuum even at zero temperature. The entire physics of Bag model is within 
this type of framework. Inside of the bag is in the
deconfining vacuum which is metastable with higher potential energy
(the bag constant). The region outside the bag is in the confining 
vacuum with zero potential energy, with a wall separating the two regions.
This is entirely consistent with the picture of a first order 
deconfinement-confinement phase transition at finite temperature where
one usually writes down a finite temperature effective potential as a 
function of an appropriate order parameter. The expectation value 
of the Polyakov line $<L>$ has been used to describe the physics of this 
phase transition at finite temperature \cite{poly,psrsk}, 
where $<L>$ is given by,

\begin{equation}
<L> = {1 \over N_c} <tr P e^{i \int_0^\beta A_0 d\tau}>,
\end{equation}

\noindent where $N_c$ is the number of colors, $\beta = 1/T$, $P$
denotes path ordering, and $A_0$ is the time component of the vector
potential. $<L>$ vanishes in the confining 
phase, while it is non-zero in the deconfining phase, breaking the Z(3) 
symmetry spontaneously \cite{poly,psrsk}.
 
 On the other hand, at $T = 0$, the physics of bag model can be captured by
postulating a color dielectric field $\chi$ with the following form of
the effective potential for $\chi$ \cite{bag1,bag2},

\begin{equation}
V(\chi) = {m_{gb}^2 \over 2} \chi^2 [1 - 2(1 - {2 \over \alpha}){\chi \over
\sigma_v} + (1 - {3 \over \alpha}){\chi^2 \over \sigma_v^2}] .
\end{equation}

\noindent with corresponding Lagrangian density being ${\it L} = {1 \over 2}
(\partial_\mu \chi)^2 - V(\chi)$. Here, $\sigma_v = 
\sqrt{2\alpha B/m_{gb}^2}$, $m_{gb}$ is the glueball mass and 
$B$ is the bag constant. We take $\alpha = 24$, $B^{1/4} = 122.3$ MeV, 
and $m_{gb}$ = 978.6 MeV. These are the parameter values used in 
ref.\cite{bag1}, and we use these as a sample. The relevance of these
parameters for us is in determining the values of surface tension $\sigma$
and false vacuum energy density $\rho$ of the deconfining bubble. We will 
present results for a wide range of $\sigma$ and $\rho$.

The plot of $V(\chi)$ is shown in Fig.1. $\chi = 0$ 
in the confining region outside the bag, while it has a non-zero value in 
the deconfining region inside the bag. $\chi$ thus has the same physical 
behavior as $<L>$, the expectation value of the Polyakov
line $L$. It would seem rather superfluous to have two entirely different
characterizations of same physical phenomenon, i.e. two different phases
of the same system, separated by an interface. One would like to think
of the color dielectric field $\chi$ as capturing the physics of $<L>$. With 
this identification, one has a single, unifying picture where the effective 
potential for a strongly interacting system is always of the form given 
in Eq.(3), with $\chi$ being interpreted as either the color dielectric 
field of the bag model, or the expectation value of the Polyakov line $L$
(neglecting the symmetry aspects of the order parameter). 
At finite temperature, the coefficients in $V(\chi)$ will become temperature 
dependent which will change the values of surface tension of the interface, 
latent heat etc. compared to the values obtained from the parameter values 
given above. The important thing is that the barrier between the deconfining 
vacuum and the confining vacuum must survive even at $T = 0$ if the physics 
of the bag model has to be preserved. We mention that the forms of effective 
potential such as the one used in ref.\cite{tph3} do not capture this physics 
of deconfinement-confinement phase transition. For example, in 
ref.\cite{tph3}, an effective potential has been used with a $T \phi^3$ 
term (with $\phi$ being the order parameter for D-C phase transition) to 
model the first order nature of the transition. Thus in this model, barrier
between the deconfining vacuum and the confining vacuum disappears below
certain temperature, which is not in the spirit of the physics of bag model.
For the same reason we do not use the parameterization used in ref.
\cite{psrsk} as that model also is appropriate for describing physics
near and above $T_c$.

 The final picture of our model can now be presented. We start with the
initial time $\tau_i$, with dense partons  filling the collision region.
This parton region will correspond to the region between the two nuclei,
receding after overlap, with transverse radius of the region being equal 
to $R_A$, the nuclear radius. Much of the collision energy will be carried 
by partons in the fragmentation region. We are interested in the total 
energy available to the expanding parton system in the central region, 
which will eventually lead to the formation of the false vacuum bubble. 
For this, we take the total energy $E_{tot}$ to be,

\begin{equation}
E_{tot} = \epsilon_i \pi R_A^2 {\Delta z}_i , 
\end{equation}

\noindent where $\epsilon_i$ is the initial energy density expected in
the collision (as given in Eq.(1)). $R_A \simeq 1.1 A^{1/3}$ fm, and 
${\Delta z}_i$ is the initial thickness of the central region. For central 
region near $z = 0$, ${\Delta z}_i$ corresponding to a rapidity interval 
$\Delta Y$ at the initial time $\tau_i$ can be taken as ${\Delta z}_i = 
\tau_i {\Delta Y}$ (see, ref.\cite{scl2}). Estimates for $\tau_i$, initial 
QGP formation time, vary from $\tau \sim 0.15$ fm (for LHC) to $\tau \sim 
0.22$ fm for RHIC. In view of various uncertainties \cite{scl1,scl2,lhc} 
in the estimates of $\epsilon_i$, and $\tau_i$, (as well as in determining
the relation between the plasma formation time $\tau_i$ and the initial
longitudinal extent of the plasma),  we will present 
results for several values of ${\Delta z}_i$. We will take ${\Delta z}_i$
= 0.15 fm for LHC, and equal to 0.22 fm for RHIC, and an optimistic value 
${\Delta z}_i$ = 1 fm for both LHC and RHIC which will correspond to a large 
value of $E_{tot}$. It is important to mention that our results only depend 
on the value of $E_{tot}$. Thus, large value of $E_{tot}$ may arise from 
uncertainties in $\epsilon_i$, or $\Delta z_i$. For Pb-Pb collision at LHC 
with $\sqrt{s} = 5.5$ TeV, expected initial energy density in the central 
region is about 1.3 TeV/$fm^3$. This gives $E_{tot} \simeq$ 25, and 175 TeV 
corresponding to ${\Delta z}_i$ = 0.15,  and 1 fm respectively. For Au-Au
collision at RHIC with $\sqrt{s}$ = 200 GeV, the energy is much smaller and
we find $E_{tot} \simeq$ 3, and 10 TeV for ${\Delta z}_i$ = 0.22, and 1 fm 
respectively.

   This parton system then expands and cools (for non-equilibrium expansion,
we say that the energy density of parton system decreases as temperature
may not be defined). Important thing is that at the initial time $\tau_i$,
the parton system must be in the deconfining phase (either due to
temperature $> T_c$, or due to non-thermal transition with extremely high 
energy density of partons). The region outside the initial parton system 
being in the confining phase, there must be an interface separating the 
two regions. Fig.2a shows such a situation. 

 Initial expansion will be longitudinal for proper times $\tau < R_A$, and 
will become three-dimensional expansion for larger times. It is expected 
that energy density (or the temperature) still remains sufficiently high so 
that by the time expansion becomes three-dimensional, the parton system 
still remains in the deconfining phase. Fig. 2b denotes such a stage of the
central parton system at $\tau \simeq R_A$. (We mention here that it is not
crucial for our model whether the fragmentation region is inside this region
shown in Fig.2b, or falls outside it.)  For subsequent expansion, we
assume (as explained in detail above) that a shell structure starts developing.
For times much larger than $R_A$, one expects an almost complete depletion of 
partons in the central region \cite{shel}, with all partons piled up near
the shell boundary. Fig. 2c denotes this intermediate stage. (We can take the 
proper time $\tau$ as measured in a frame at rest at some average point in 
the middle of the shell. Velocity there will not be very close to $c$, 
so $\tau$ will not be much different from the time $t$ in the center of mass 
frame. In any case, the  3-dimensional expansion is not expected to be
ultra-relativistic, so at this stage, one can use $\tau$ and lab time $t$
interchangeably. We also neglect any Lorentz contraction in the
thickness of the parton shell.) The stage shown in Fig.2c is the non-trivial 
part of our model. Here, the entire region inside the sphere of radius 
$\simeq ct$ is taken to be in the deconfining vacuum. There is then the 
interface at the boundary. The partons accumulate near the boundary of 
the shell from inside. The thickness of this region containing the 
partons will be expected to be of order of  about $2 \times R_A$ as 
that is the total extent  of the region at the time at which three 
dimensional expansion begins.

The region with $r < ct - 2R_A$ has no partons, but is still in the 
deconfining vacuum. This somewhat unconventional picture can be further
justified as follows. Starting from a stage as shown in Fig.2b, where the 
entire interior region was in the deconfining vacuum, one reaches the stage 
shown in Fig.2c by parton expansion. The inner region (devoid of partons) 
can turn into the confining vacuum only if there is nucleation of bubbles of 
confining vacuum in that region, which upon coalescing will convert the inner
region into the confining vacuum. This could happen if the inner region cooled
to $T = 0$ (with zero parton density, assuming zero baryon number in the 
central region) maintaining thermal equilibrium. However, thermal nucleation 
of bubbles could not happen since we are assuming that freezeout occurred
before the hadronization transition (or, due to large time scales of bubble 
nucleation \cite{kpst}).  Even if the system did not go through finite 
temperature bubble nucleation when temperature decreased below $T_c$, 
the inside region can still be in the 
confining vacuum if quantum nucleation of confining vacuum bubbles (at $T = 
0$) could happen. As we will discuss below, the action $S_0$ of such a bubble 
is about 100 in natural units. The probability of nucleation of such a bubble 
is proportional to $e^{-S_0}$, and hence is completely negligible for the
relevant time scales, allowing for reasonable dimensional estimates for the 
pre-factor for the nucleation probability. (As in our model a significant
barrier separating the two vacua remains even at $T = 0$, there is no
possibility of the phase transition occurring via spinodal decomposition.)
Thus, we conclude that the inner region of the shell must correspond to 
$T = 0$ supercooled deconfining vacuum.

 Eventually, as parton shell keeps expanding, its energy density falls below
a critical value so that hadronization must take place. Partons will then form
hadrons as they leave the spherical region bounded by the interface. This
is the stage when the interface will stop moving outward, and will start 
shrinking. As the interface passes through the parton system, it converts 
partons into hadrons, as shown in Fig.2d. (Alternatively, hadronization may
happen at various places in the parton shell. This will also amount to
an effective shrinking of the interface.) The motion of the interface will be 
dissipative at this stage due to its interaction with partons. Typical 
velocity of the interface, therefore, will be less than the speed of 
sound \cite{vel,tph3,kpst}. However, as all (or most) of the partons
hadronize, the interface 
will continue to move inward due to negative pressure of the metastable
vacuum. Now there are no partons to impede the motion of the interface.
What one is left with is a spherical bubble of pure false vacuum. 
To obtain the profile of the wall of this {\it false vacuum} bubble, we 
first obtain the profile of the {\it true vacuum} bubble as follows.
Given $V(\chi)$ in Eq.(3), one can obtain the instanton solution for the
tunneling through the barrier from $\chi \ne 0$ metastable vacuum to the
$\chi = 0$ true vacuum. This is given by the solution of the following 
equation \cite{inst},

\begin{equation}
{d^2 \chi \over dr^2} + {\kappa \over r} {d \chi \over dr} -
V^\prime(\chi) = 0 ,
\end{equation}

\noindent where $V(\chi)$ is the effective potential in Eq.(3) and $r$ is the 
radial coordinate in the Euclidean space. In the Minkowski space initial 
profile for this true vacuum bubble is obtained by putting $t=0$ in the 
solution of the above equation. Parameter $\kappa$ = 3 for quantum nucleation 
of bubble (at $T = 0$), while $\kappa$ = 2 for thermal nucleation of bubble 
at finite temperature. Initially the parton region had large temperature, so 
the profile of the interface at that stage would be given by the finite 
temperature bubble (i.e. $\kappa = 2$). This will be true up to the stage 
shown in Fig.2b. After freezeout, or once the partons leave the shell in the 
form of hadrons as in Fig.2e, the profile of the interface  will be given by 
$T = 0$ bubble, that is $\kappa = 3$ case. We have solved Eq.(5) using a fourth
order Runge Kutta algorithm for the effective potential as given in Eq.(3). 
The numerical technique is the same as used earlier for standard false vacuum 
decay \cite{bbl} with the obvious difference that now the false vacuum occurs 
at non-zero value of $\chi$ while the true vacuum occurs at $\chi = 0$. Thus 
the search for the bounce solution has to be done differently in the present
case. In Fig.3a we have shown the solution of the $T = 0$ quantum bubble
of true vacuum.  The radius of this bubble is about 5 fm. (The finite 
temperature bubble has a smaller radius $\simeq$ 3.8 fm.) We have calculated
the action for the true vacuum bubble shown in Fig.3a and we find it to
be about 100 in natural units. As mentioned above, this large value of
the action suppresses the nucleation of true vacuum bubbles in the the 
central deconfined (T = 0) region, so that this region converts to the 
confining vacuum only via collapse of the spherical interface shown in
Fig.2e.

Fig.3a shows the bubble of confining vacuum embedded in the deconfining 
vacuum (as obtained by the instanton solution of Eq.(5)). However, the 
experimental situation we have discussed above is exactly the reverse. 
As shown in Fig.2e, we have a large bubble of deconfining vacuum which is 
embedded in the confining vacuum. Such a bubble cannot be obtained as a 
classical solution of Eq.(5). The situation shown in Fig.2 arises because 
of changing temperature, and is similar to the false vacuum bubble
formation due to coalescence of true vacuum bubbles as discussed in the 
context of quark-hadron transition in the early universe \cite{wtn}. 
Main difference between our case and the case discussed in ref.\cite{wtn} 
(see, also ref.\cite{kpst}) is that there the motion of bubble wall 
remains dissipative due to presence of QGP in the interior of the bubble. 
In contrast, in our case, the special geometry of collision gives the 
shell like structure which leads to a pure false vacuum bubble with no 
QGP inside. The motion of the interface, therefore, will not be 
dissipative. Fig.3b shows this false vacuum bubble
which is simply obtained by inverting the profile of the bubble as given in
Fig.3a. This is a somewhat approximate way of generating the appropriate 
bubble profile. However, our interest is only in the values of surface tension
$\sigma$ of the interface and the false vacuum energy density $\rho$. (Note
that $\rho$ is the same as the bag constant $B$. We use a different notation 
for it due to the method by which we calculate it by generating the confining 
vacuum bubble and then inverting it.) We will present results for a range of 
values of these parameters, with the profile given in Fig.3b corresponding to 
one set of these values. Note that the radius of this deconfining vacuum 
bubble (Fig.3b) is not obtained from equation of motion. In contrast, radius 
of the confining vacuum bubble of Fig.3a was fixed by the solution of Eq.(5). 
Inverted profile in Fig.3b is generated only to estimate the values of $\sigma$
and $\rho$ corresponding to parameter values in Eq.(3). Its radius will be 
determined by the physics of the problem, such as total energy of collision, 
as we discuss below. In the same way, the radius of the false vacuum bubble 
in ref.\cite{wtn} is determined by separation between the nucleation sites 
of the hadronic bubbles, and the manner in which they coalesce.

 To obtain values of surface tension $\sigma$ and the false vacuum energy
density $\rho$ of the deconfining vacuum bubble, we obtain total energy $E$ 
of the false vacuum bubble by numerically integrating the energy
density, ${1 \over 2} 
(\bigtriangledown \chi)^2 + V(\chi)$, for the bubble profile in Fig.3b. 
Bubble energies are obtained in this manner for two different values of 
bubble radius. $\sigma$ and $\rho$ are then determined by using the 
following equation for the two values of bubble energy and radius,

\begin{equation}
E = 4 \pi r^2 \sigma + {4 \pi \over 3} r^3 \rho .
\end{equation}

The values of $\sigma$ and $\rho$ obtained by using this equation for two 
values of bubble radius are tested for other bubble radii and it is found 
that bubble energies are accurately reproduced. For the parameters used for
Eq.(3) in ref.\cite{bag1}, we find $\sigma = 64.8 {\rm MeV/fm}^3$ and $\rho = 
27.8 {\rm MeV/fm}^3 (\simeq (122 {\rm MeV})^4)$. Note that this value of 
$\rho$ is the same as the bag constant $B$ in Eq.(3) as it should be. This
gives us confidence that this procedure of inverting profile of confining 
vacuum bubble to get the profile of the deconfining vacuum bubble is
reasonably accurate. For comparison, we repeated this procedure for the finite 
temperature bubble, i.e. solution of Eq.(5) with $\kappa$ = 2, (pretending 
that the effective potential in Eq.(3) corresponds to the finite temperature 
case). As we mentioned, resulting bubble radius of the confining vacuum
bubble (similar to the $T = 0$ bubble in Fig.3a) is about 3.8 fm. We find
that the inverted bubble (deconfining vacuum bubble, as in Fig.3b) has
$\sigma = 64.4$ MeV/fm$^2$ and $\rho$ = 27.8 MeV/fm$^3$. These values are
essentially the same as in the case of $T = 0$ bubble. Since the only thing 
we need from the above calculations is the values of these parameters, it is 
immaterial whether we think of the interface as corresponding to the finite 
temperature bubble or to $T = 0$ bubble.

\section{Expansion and Subsequent Ultra-Relativistic Collapse of the Shell}

 We need to make one further specification about the expansion of the
parton system. If the expansion is adiabatic, then total entropy will be 
conserved. In that situation, total energy will decrease due to the work
done by the pressure of the plasma. This description is appropriate when 
expansion happens maintaining equilibrium, as in the hydrodynamical models 
of parton expansion. In the situation of free expansion, or if there are 
dissipative effects present, then expansion could be energy conserving
\cite{de0}. We have argued above that shell picture is  more reasonable in the 
case when parton system freezes out early. Thus energy conserving expansion
may be more appropriate for our model. We consider this case first. We will 
also briefly discuss results for the adiabatic expansion, and we will
see that the temperatures of the hot spot are much smaller for that
case. (Note that the only 
thing relevant to us about the parton expansion is the development of a shell
structure with central region devoid of partons, but still in the deconfined 
vacuum. We have argued that this seems natural when partons freezeout early. 
However, all of our arguments are valid if this picture can be justified 
even by taking partons to be in equilibrium, through the stages in Fig.2a-2d.) 
Assuming the energy to be conserved, the energy density of the partons in
the shell, $\epsilon(r)$, at any stage shown in Fig.2c, is determined by,

\begin{equation}
4\pi r^2 \sigma + {4\pi \over 3} (r^3 - (r - \Delta r)^3) 
\epsilon(r) + {4\pi \over 3} r^3 \rho  = E_{tot},
\end{equation} 

\noindent where $E_{tot}$ is given in Eq.(4). Here $\epsilon(r)$ denotes 
the parton energy density in the shell of thickness $\Delta r \simeq 2 
R_A$. The shell will expand to a largest size $r_{max}$ at which stage 
parton system will hadronize. We determine $r_{max}$ by taking $\epsilon(r) 
= \epsilon_c$, the critical value of energy density of partons below which 
hadronization is expected to take place. We take $\epsilon_c \simeq {\pi^2 
37 \over 30} T_c^4$, where $T_c$ is the critical temperature for the 
quark-hadron transition. $\epsilon_c$ is not to be taken necessarily as the
energy density of plasma in equilibrium (with temperature = $T_c$), as we have 
argued that the partons may be in a non-equilibrium state. We take $\epsilon_c$
as giving the scale below which partons should convert to hadrons. 
The value of $T_c$ is related to the bag constant (for an equilibrium 
transition) using Gibbs criterion of equal pressure at transition 
temperature, $T_c = (B/(g_q - g_h))^{1/4}$. Here, 
$g_q = 37\pi^2/90$ and $g_h = 3\pi^2/90$, corresponding 
to 2 massless quark flavors and 8 gluons in the QGP phase, and 3 pions in 
the hadronic phase. Again, the value of $B$ appropriate at the scale
given by $T_c$ may be different from the value of $B$ at $T = 0$.
For simplicity, we ignore any scale dependence of $B$ and use the 
same value of $B$ which equals the false vacuum energy density $\rho$, 
as also determining $T_c$ via above relation. Therefore, we present 
results for a large range of values of $B$, not limiting to only those 
values which correspond to realistic values of $T_c$. 
 
 $r_{max}$ gives the largest radius of the shell, with interface being at 
the outer boundary (i.e. at $r = r_{max}$) initially. As this is the stage 
when hadrons start forming, interface starts moving inwards. Initial motion 
of the interface is highly dissipative \cite{vel,tph3}, as long as it
traverses the region which is filled with the partons, i.e. a thickness of 
$\Delta r \simeq 2 R_A$. This stage is shown in Fig.2d. Once the interface 
has shrunk below this parton filled region, it is free to undergo unimpeded, 
relativistic collapse. The region bounded by this interface at a later stage 
(shown in Fig.2e) represents a pure false vacuum bubble with no partons 
inside. The initial radius of this pure false vacuum bubble, $r_f$ will be
approximately given by,

\begin{equation}
r_f \simeq r_{max} - R_A ,
\end{equation}

\noindent where $r_{max}$ is determined by solving Eq.(7) with $\epsilon(r)
= \epsilon_c$. In writing this expression for $r_f$, we have taken that when 
$\epsilon(r) = \epsilon_c$ is achieved during the shell expansion, the 
interface starts collapsing with almost speed of light, as the parton shell 
keeps expanding relativistically. However, due to dissipative motion of the 
wall through the parton system, velocity of the wall will be typically much
smaller than the speed of light \cite{vel}. If one takes the wall to be 
almost static, as compared to the outward velocity of the shell, then one 
will expect $r_f$ to be almost equal to $r_{max}$. Thus, the value of $r_f$ 
as given above is an underestimate. Resulting total energy of the false
vacuum bubble, and hence the final temperature of the hot spot will also
be somewhat underestimated in using the above equation. This should hopefully
compensate, to some degree, the effects of assuming all partons to be
ultra-relativistic, (e.g., if the parton distribution extends beyond the
assumed shell thickness of $2 r_{max}$, it will reduce the value of $r_f$). 

Total energy $E_f$ of this pure false vacuum bubble is, 

\begin{equation}
E_f = 4 \pi r_f^2 \sigma + {4 \pi \over 3} r_f^3 \rho .
\end{equation}

  Further evolution  of this false vacuum bubble is well understood.
It will undergo relativistic collapse. Due to surface tension, bubble
will become more spherical as it collapses. This is important as there are
various factors because of which the initial shape of the bubble wall
may not be spherical. First, the collision geometry itself will not be
expected to give rise to completely spherical structure. Secondly, the 
interface motion through the parton shell itself may not be very isotropic. 
However, during the free collapse of the interface, the shape should become
more spherical due to surface tension. Any remaining asphericity will
ultimately affect the final radius to which the bubble can collapse before
the bubble walls decay via collision. For simplicity, we will assume that the
interface collapses maintaining spherical symmetry. During bubble collapse, 
the potential energy of the false vacuum simply gets converted into
the kinetic energy of the collapsing interface \cite{inst}. This is where
our model crucially differs from the collapsing QGP bubbles previously
discussed in the context of quark-hadron transition \cite{wtn,kpst}. There,
interface always moves through a QGP system filling the interior of the
bubble, which impedes the motion of the interface. The false vacuum 
energy there gets converted to the heat which raises the temperature of
the plasma \cite{wtn,kpst}. In contrast, there are no partons in the interior
of the deconfining vacuum bubble in our model. The entire false vacuum 
energy thus converts to the kinetic energy of the interface. The collapsing
interface quickly becomes ultra-relativistic, with extremely large
Lorentz contraction factor, as indicated by thinner interface in Fig.2e.

 At any stage during the collapse of the shell, the value of the Lorentz
$\gamma$ factor primarily depends on the initial radius $r_f$ of the shell,
and the value of $\rho$ (i.e. $B$). To give an idea of how rapidly the bubble
wall Lorentz contracts, we give values of $\gamma$ at the stage when the
bubble has collapsed to a radius of 1 fm. For Pb-Pb collision with
$\sqrt{s}$ = 5.5 TeV, and with ${\Delta z}_i$ = 1 fm, we find that $r_f$
varies from 20 to 90 fm as $B^{1/4}$ is reduced from 240 MeV to 120 MeV (as
we will see below). Value of $\gamma$, when bubble has collapsed to a radius 
of 1 fm, ranges from $3 \times 10^4$ to $8 \times 10^4$. For same parameters, 
but with ${\Delta z}_i$ = 0.15 fm, $\gamma$ ranges from 2000 to 7000. At RHIC 
energy, with $\sqrt{s}$ = 200 GeV for Au-Au collision (with ${\Delta z}_i$ 
= 1 fm), $r_f$ ranges from 5 fm to 35 fm as $B^{1/4}$ is reduced from 250 MeV 
to 100 MeV. Resulting $\gamma$ factor (again, when bubble has collapsed to 1 
fm radius) ranges from about 500 to 4000. For this case, by the stage when 
bubbles radius has decreased by another order of magnitude, i.e. to a value 
of 0.1 fm, $\gamma$ ranges from about $5 \times 10^4$ to 
$3 \times 10^5$.
 
  The thickness of the interface initially (i.e. at stages shown in Fig.2a-2d),
is about 1 fm. We see that by the time bubble collapses to a radius of
about 1 fm, the thickness of the interface is much smaller, about $10^{-3}$
fm to $10^{-5}$ fm. Interface thickness reduces to about $10^{-5}$ fm to
$10^{-7}$ fm by the time the bubble collapses to a radius of 0.1 fm. Again,
it is unconventional to talk about such large length contraction factors
in the context of heavy-ion collisions. We know that due to virtual partons
at small x (the so called {\it wee partons}), nucleus thickness does not
Lorentz contract below about 1 fm even at ultra-high energies \cite{wee0}
(see, also, ref. \cite{wee1}). However, this limiting Lorentz contraction 
arises due to virtual particle production in the color field of 
the partons. In our case, by the time the interface contracts 
below the partonic shell, there will be no such virtual partons inside
the bubble. Thus, there is no compelling reason to believe that the 
Lorentz contraction of the interface of this pure false vacuum bubble 
should be limited by the typical QCD scale of 1 fm. Basically, the radial 
profile of the interface represents the kink solution (Eq.(5)) which can be 
Lorentz boosted to any velocity in the absence of interactions with partons. 
Even with quantum fluctuations on the background of this ultra-relativistic
bubble wall, there is no reason to expect that these fluctuations will
put a stop to the highly Lorentz boosted bubble wall. Though some of
the wall energy may get dissipated due to these effects, bubble collapse
may still continue to sizes much smaller than 1 fm as long as walls remain
ultra-relativistic.

 This extreme Lorentz contraction of the bubble wall has following important
consequence. Normally, the process of the collapse of a bubble wall will
stop when its radius is of the order of the thickness of the wall. (See, for 
example, the numerical study of collapsing domain walls in ref.\cite{dmnwl}.
Though, for ultra-relativistic collapse numerical errors can build up and 
one needs more sophisticated numerical techniques, see ref.\cite{chptnk}.)   
Without significant Lorentz contraction, we would expect the bubble collapse
to halt at the stage when bubble radius is of the order of 1 fm, as in the 
standard treatments of collapse of QGP bubbles in a quark-hadron transition
\cite{wtn}. Then bubble walls (of different bubble portions) will collide and
the energy of walls will be released in the form of particle production.
However, in our case the bubble wall undergoes large Lorentz contraction
even when bubble radius is about 1 fm (in the center of mass frame). As 
compared to the wall thickness, bubble radius is about $10^3$ to $10^5$ 
times larger at this stage. Thus, there is no reason to expect that bubble 
collapse will halt at this stage. In fact it is easy to see (using the fact
that as bubble collapses, false vacuum energy gets converted to the kinetic
energy of the bubble wall) that bubble wall thickness decreases much faster 
(due to Lorentz contraction) compared to the bubble radius, as bubble collapse 
proceeds. Of course, eventually this type of {\it classical} evolution of 
bubble must stop, at least by the stage when the net size of the system has 
become smaller than that allowed by the uncertainty relation. The net energy
$E_f$ we start with (of the false vacuum bubble at the initial stage when free
collapse of the interface commences), is much larger than a TeV. Thus, when 
collapsing bubble size becomes smaller than $E_f^{-1}$ (say a TeV$^{-1}$), 
it seems reasonable that bubble collapse may halt. Entire energy $E_f$ of 
the initial pure false vacuum bubble, which was subsequently converted into 
the kinetic energy of ultra-relativistic walls, may thus get converted to 
particles in a region of size less than $(E_f)^{-1}$.  
(It will be interesting to work out the exact conditions when 
bubble collapse will halt. Even if the bubble radius shrinks only to a value 
of say 0.1 fm, one still gets a hot spot, though with a lower temperature.
It is also possible that in some situations, bubble collapse may develop 
strong anisotropies. In that case the bubble may break into smaller bubbles.) 

 It is important to realize that the above picture only requires that
most, or a significant fraction, of the bubble wall energy (after bubble
has collapsed to size of order 1 fm) is not dissipated away subsequently 
until different portions of the (Lorentz contracted) bubble wall collide 
with each other. This is because essentially all of the energy of the
initial false vacuum bubble is already stored in the bubble walls by the
time its radius shrinks to a value of order 1 fm (for initial
radius much larger than 1 fm). Even with quantum fluctuations on the bubble 
wall background, there is no reason to expect that these fluctuations will 
tend to put an early stop on (or significantly dissipate the kinetic
energy of) the highly relativistic bubble wall.  For example, even for 
the case of nucleus-nucleus collision, interactions of wee partons do not 
stop the nuclei from overlapping (or going through) at sufficiently large 
collision energies. Though, we again emphasize that a nucleus, filled with 
partons, is a very different object than a {\it pure false vacuum bubble}. 
Thus, the arguments based on wee partons made for nucleus \cite{wee0} do 
not extend in a natural way to the Lorentz contracted bubble wall. After all,
one does not even know what is the correct field describing such a 
wall (apart from various possible order parameters), let alone the
detailed nature of quantum fluctuations  above the bubble wall separating 
deconfined and confined vacua of QCD (at zero temperature) and their 
Lorentz transformation properties. From all this, we would like to 
conclude that there seems a genuine possibility that bubble collapse
may continue, preserving most of its energy, down to sizes much smaller
than 1 fm. As we will see below, this possibility leads to very
interesting implications.

  The final conclusion is that the entire energy $E_f$ of the false vacuum 
bubble will be converted to a dense system of partons, contained in an 
extremely tiny region which can be as small as few TeV$^{-1}$ to begin 
with (or even smaller). The energy of 
this bubble, $E_f$, is a fraction of the total energy 
of the initial parton system $E_{tot}$ as given in Eq.(4). For $A = 200$,
and with ${\Delta z}_i$ = 1 fm,  this fraction ranges from 15\% to 40\% for 
$\sqrt{s} = 5.5$ TeV, and 20\% to 50\% for $\sqrt{s}$ = 30 TeV, as $B^{1/4}$ 
is decreased from 240 MeV to 120 MeV. For ${\Delta z}_i$ = 0.15 fm, these 
fractions range from 5\% to 20\%, and 10\% to 35\% for the two values
of $\sqrt{s}$ respectively. For RHIC, with $\sqrt{s} = 200$ GeV, and 
with ${\Delta z}_i$ = 1 fm, this fraction ranges from about 5\% to 
25\%, as $B^{1/4}$ is decreased from 220 MeV to 100 MeV. We 
mention that here, as well as in later sections, for LHC, we will be 
considering a range of values of $\sqrt{s}$, including very large values 
such as 30 TeV (for Pb-Pb collision). This is with the idea that any 
possibility of observing new physics 
at such large energies, should provide strong motivation
for going for such large (or even larger) values of $\sqrt{s}$.

\section{Equilibration of decay products of bubble wall}

 Most of the energy of the initial parton system escapes out in the form of 
hadrons, and as we have seen above,  only a fraction is left behind in 
the form of the vacuum energy inside the bubble. However, the important 
thing is that all this energy gets focused into a very tiny region due to 
ultra-relativistic collapse of the bubble wall. The resulting energy
density can be extremely high. This parton system will then expand and in 
that process thermalize. For consistency of thermodynamic equilibrium,
we must have a region of size at least of the order of the mean free
path $r_{eq}$ of the relevant degrees of freedom. For the relevant
values of $r_{eq}$ and the temperature, we find that due to small size 
of the hot spot, only strongly interacting particles can be in equilibrium.
In naive perturbation theory, one would use interaction rate
$\Gamma \sim \alpha_s^2 T$. With the value of $r_{eq} \sim (\alpha_s^2 
T)^{-1}$, the resulting temperatures of the hot spot are rather low. 
However, as has been discussed in the literature \cite{qcdft1,qcdft2}, 
for QCD at finite temperature, there are serious problems such as 
infrared divergence and gauge dependence of results. An improved
perturbation theory has been introduced by Braaten and Pisarski 
\cite{qcdft1}. In this hard thermal loop re-summation technique, the 
relevant scattering cross-section rates are given as $\sim \alpha_s T$ for 
quark-quark scattering and $\sim 2 \alpha_s T$ for gluon-gluon 
scattering \cite{qcdft2}. With these results
in view \cite{qcdft2}, we will estimate the resulting temperature $T$
of the hot spot when its size is equal to,

\begin{equation}
 r_{eq} \simeq (2.2 \alpha_s T)^{-1}
\end{equation}

and,

\begin{equation}
 r_{eq} \simeq (\alpha_s T)^{-1}
\end{equation}

 First value of $r_{eq}$ corresponds to the large interaction rate of
gluons (say, for pure gluonic case \cite{qcdft2}) and is relevant when 
only gluons equilibrate (size of the region being too small for quarks 
to effectively scatter). The second case (Eq.(11)) corresponds to the 
equilibration of the combined quark-gluon system.

  Note, however, that the dynamics of equilibration of a rapidly
evolving parton system is not too well understood. For example,
estimates for equilibration time of partons at LHC show 
\cite{klauss} that
the parton distributions approach equilibrium distributions within
a duration of about 0.15 fm/c (with resulting QGP temperature being
about 1 GeV). This time scale is too short compared to either of the
estimates of $r_{eq}$ given above. Even for g-g scattering the
scattering rate given above will imply a time scale of about 0.5 fm 
(with $\alpha_s \sim 0.2$ for $T \sim 1$ GeV), for q-q scattering one should 
have expected a time scale of about 1 fm/c. Similarly, for RHIC, the 
estimated equilibration time\cite{klauss} of 0.22 fm/c (with resulting
QGP temperature of about 500 MeV) is still shorter than what one will
get from above estimates of scattering rates. Keeping this in mind,
we will also allow for the possibility that the decay products of
bubble wall may also thermalize within a shorter time duration which
we take to be 0.15 fm/c (this should be reasonable for hot spot 
temperature of about 1 GeV). Thus, along with the values of $r_{eq}$
as given above by Eq.(10)-(11), we also consider the following value.

\begin{equation}
 r_{eq} \simeq 0.15 {\rm ~fm}
\end{equation}

 It is clear that the use of constant $r_{eq}$ can only be hoped to
be a reasonable approximation over a limited temperature range.
One may trust this value when temperature of the hot spot is
obtained to be in the range of a couple of GeV. 

An important point to note here is that in general the decay
products of the bubble wall can include other particles as well,
e.g. leptons, Higgs bosons etc. However, the electroweak coupling
being much smaller, these degrees of freedom will take much longer
to equilibrate, with effective temperature of the hot spot being
very low (or, by that time the decay products may even freezeout).
This means that only a fraction {\it f} of the total energy of the false
vacuum bubble will be available in terms of quarks and gluons,
remaining energy being carried by particles streaming out of the 
dense parton region. This is similar to the streaming out of direct
photons and leptons from the early stages of parton production in 
relativistic heavy-ion collisions. Since the bubble wall separates the 
two phases of QCD, it is possible that its decay products
will predominantly be quarks and gluons due to their larger
cross-sections. Thus for estimating the
maximum temperature of the equilibrated hot spot we will first assume 
that the entire energy of the false vacuum bubble goes in creating a
thermal system of quarks and gluons.

 With these arguments, the resulting temperature $T_f$ of the hot spot 
at this stage can be determined by the following equation, 

\begin{equation}
 {g_* \pi^2 \over 30} T_f^4 {4\pi \over 3} r_{eq}^3 + 4\pi r_{eq}^2 
\sigma + {4\pi \over 3} r_{eq}^3 \rho ~=~ E_f
\end{equation}

 Here, $g_* = 37$ is the number of quark-gluon degrees of freedom.
$\sigma$ and $\rho$ are the values of interface tension and false
vacuum energy density as used earlier in Eq.(7) and Eq.(9). For the 
size of the region $r_{eq}$ we will take various values as given by 
Eqs.(10)-(12). We use the following expression \cite{alpha} for $\alpha_s$ 
as a function of temperature in Eqs.(10)-(11),

\begin{equation}
 \alpha_s(T) = {6\pi \over (33 - 2 N_f) ln({8T \over T_c})}
\end{equation}

\noindent where $T_c$ is the critical temperature, and we use number
of flavors $N_f = 2$. With this expression for 
$\alpha_s$, Eq.(13) is solved numerically to get $T_f$ self consistently.

  Note that when taking $r_{eq}$ from Eq.(10), we are assuming that
due to larger interaction rate, only gluons may thermalize, with 
quarks remaining out of equilibrium (within the region of size
$r_{eq}$). In such a case $g_*$ should be changed from 37 to 16, while
at the same time $E_f$ in the right hand side of Eq.(13) should be
reduced by a factor ${16 \over 37}$ (assuming equipartition of energy) 
as only this fraction of total
bubble energy will thermalize. It turns out that this modification
does not affect any results, the two factors effectively compensating
for each other (i.e. other terms in Eq.(13) remain subdominant). We
will thus not worry about this modification of the above equation and
will continue to use Eq.(13) for all values of $r_{eq}$ from
Eqs.(10)-(12).

\section{Results}

 We have obtained the value of $T_f$ for  a range of values of bag constant 
$B$, and for different values of $A$ and initial energy density $\epsilon_i$ 
(which can be related to $\sqrt{s}$ using Eq(1)). The value of surface tension 
of the wall $\sigma$ is taken to be 64 MeV$/fm^2$ (for Eqs.(7),(9)), as 
determined using Eq.(6). We find that our results are rather insensitive to 
the value of $\sigma$. Changing $\sigma$ to 1 MeV/$fm^3$ leads to virtually no
change in $T_f$ and $r_f$. Note that this means that our quantitative
results may be valid even if the barrier height between the
deconfining and confining vacua is very small, as long as shell
structure can be justified. Increasing $\sigma$ to even unreasonably 
large value of 300 MeV/$fm^3$ increases $T_f$ by about 
6 \%, and decreases $r_f$ by about 10 \%. 

\subsection{LHC}
 
 Here, we present results for energies suitable for LHC. First we consider 
the case when only gluons equilibrate, so $r_{eq} = 2.2 \alpha_s T$ 
(Eq.(10)). The physics here will be similar to the {\it two
stage} model discussed in ref. \cite{twostg} where gluons first
thermalize to give higher temperature of the plasma, while quark
equilibration takes a longer time resulting in lower temperature of
the combined system. Here also we see that the largest values of $T_f$ 
are obtained for this case. In Fig.4 we have plotted $T_f$ (in GeV) 
vs. $B^{1/4}$ (in MeV) for the case when $\Delta z_i$ (in Eq.(4)) is 
1 fm. To give an idea how $T_f$ varies with $A$, we give plots for
three different values of $A$. Solid, dotted, and dashed curves
correspond to $A$ = 200, 100, and 50 respectively. For subsequent figures,
we  will give plots only for the case $A = 200$. In Fig.4, plots
have  been given for $\sqrt{s} = 30, 15, 5.5$ TeV, which 
can be translated to the values of $\epsilon_i \simeq 4.9, 2.8, 1.3 ~ {\rm 
TeV/fm^3}$, respectively (for $A = 200$). Fig.4 also shows plots
of the radius $r_f$ of largest pure false vacuum bubble for $\sqrt{s}$ 
= 5.5 TeV. Note, the shell thickness $\Delta r$ for these cases is of 
order 2$R_A \simeq 13$ fm.

 Values of $T_f$ are large in Fig.4 due to large interaction rate of 
gluons leading to rapid thermalization. 
This will be the highest temperature of an equilibrated system
which can be expected in our model. Subsequently, quarks will also
thermalize and the appropriate value of $r_{eq}$ will be given by
Eq.(11). The resulting values of $T_f$ for the combined quark-gluon
system are smaller as shown by the dotted plots in Fig.5. In this
figure we also show plots (shown by dashed curves) for the case 
when $r_{eq}$ is taken to have a fixed value equal to 0.15 fm. 
As we mentioned earlier in Sec.V, parton cascade simulations
suggest that various particle distributions approach equilibrium
distributions within a time duration which may be as short as 0.15 fm/c
for LHC with the associated QGP temperature of about 1 GeV. This time 
scale is much shorter than what one gets from the interaction rate in Eq.(11).
Even the gluonic interaction rate (Eq.(10)) gives a significantly longer
time scale. Thus we show plots for $r_{eq} = 0.15$ fm to allow for this 
possibility. As we mentioned in Sec.V, using a
fixed value of $r_{eq}$ means that the plots are not going to be
reliable for a wide range of the temperature $T_f$ of the hot spot. 
However, for the values of $T_f$ around few GeV the plots may lead
to reasonable estimates. 

 Plots in Figs.4,5 are with value of $\Delta z_i = 1$ fm in Eq.(4). 
In Fig.6 we show plots for the case when $\Delta z_i = 0.15$ fm. 
Resulting temperatures are much smaller now.
As can be clearly seen from these plots, the temperature of the hot
spot can easily reach well above 1 GeV, which is the expected initial
temperature of the QGP here. Such a hot spot can lead to clean signals.
Even for a reasonably conservative set of parameters, $r_{eq} = 0.15$
fm, $\Delta z_i = 0.15$ fm, and for the values of $B^{1/4}$ consistent
with the critical temperature of about 170 MeV, the value of $T_f$ can
easily reach around 3 GeV (for $A = 200$, and with $\sqrt{s} =$ 5.5 TeV). 

\subsection{RHIC}

 For RHIC, with $\sqrt{s}$ = 200 GeV, for Au-Au collision, we find that 
$T_f$ is very small and rarely reaches above 1 GeV. In Fig.7 we
give plots for two choices, $\Delta z_i$ = 1 fm and = 0.22 fm.
For $\Delta z_i = 1$ fm case, solid and dashed plots 
correspond to $r_{eq}$ given by Eq.(10) and Eq.(12) respectively. No
solution is found for the case when $r_{eq} = \alpha_s T$ (Eq.(11))
for reasonable values of $B^{1/4}$. For $\Delta z_i = 0.22$ fm,
solutions for reasonable values of $B^{1/4}$ are found only 
for the case when $r_{eq} = 0.15$ fm,
as shown by the dashed plot.  (As the temperature reaches 
about a GeV for this case, $r_{eq} = 0.15$ fm, which is the expected
equilibration time at LHC, may be appropriate to use here).
$A = 200$ for all these plots.

We have repeated the entire analysis for the case when entropy is conserved 
during parton expansion. In this case values of $T_f$ are much smaller. We
quote some values here to give an idea of the typical range of values of 
$T_f$ for this case for various parameter values (with $A = 200$). As
the largest values of $T_f$ are obtained for the case when only gluons
equilibrate, we give here numbers for the case when $r_{eq} = 2.2 
\alpha_s T$. For  $\sqrt{s}$ = 30 TeV and with ${\Delta z}_i$ = 1 fm, 
we get $T_f$ varying from 6.2 GeV to about 5.3 GeV as $B^{1/4}$ 
increases from 100 MeV to 240 MeV. (The range of $T_f$ becomes from
3.9 GeV to 3.3 GeV for $A = 100$.) For $\sqrt{s}$ = 15 TeV, the 
corresponding values of $T_f$ range from 4.2 GeV to about 3.5 GeV.
For $\sqrt{s}$ = 5.5 TeV, $T_f$ ranges from 2.4 GeV to 2 GeV for the
same range of $B^{1/4}$. For ${\Delta z}_i$ = 0.15 fm, $T_f$ is much 
smaller, ranging from 1.2 GeV to 0.8 GeV even with $\sqrt{s}$ = 30 TeV. We 
mention again, as discussed above, due to non-equilibrium conditions, 
energy conserving  expansion may be more appropriate in our case. 

\subsection{Possibility of non-thermal restoration of electroweak
  symmetry?}

 So far we have presented results estimating the temperature of the hot
spot, assuming the equilibration of the decay products (or a fraction of
them) resulting from the bubble wall. 
As the collapsing bubble concentrates a very large energy in a very
tiny region, the resulting initial parton system will be extremely
dense. The decay of ultra-relativistic bubble walls should lead to
rapid particle production. When wall energy is sufficiently high
(say for LHC case) then decay products may also consist of a sizeable 
density of the Standard
model Higgs particles among other decay products. This type of picture 
reminds one of the rapid particle production at the end of
inflationary phase in the early Universe where the inflaton field 
decays into particles. A very interesting possibility which has been 
proposed in that context \cite{rsnc} is that, during the early
stages of particle production,  even when the system
remains in a non-equilibrium state, one may still get restoration of
symmetry due to modification of the effective potential from large
population of particle modes resulting from the decay of the inflaton.

 This raises the question whether a similar phenomenon can also
occur within the context of our model. The decay of
ultra-relativistic bubble walls will lead to rapid population of
various particle modes (including the scalar particles). It may then
be possible for the {\it electroweak} effective potential to be
modified, leading to effective restoration of electroweak symmetry,
even if the whole system remains in a non-equilibrium state. 
In fact, even for the early evolution of the quark-gluon system
resulting from collision of the two nuclei, one may be able to study
this possibility and see if, for example, chiral symmetry may get
restored at a very early stage when the whole parton system is still 
in a non-equilibrium state. We hope to check both these possibilities 
more carefully in future. For our present purpose, we will simply
assume that such a non-thermal restoration of electroweak symmetry can indeed 
happen, and present the estimates corresponding to such a scenario. 

  It is fair to say that at this stage this remains a speculation.
Thus the results of this subsection should only be
taken as representing many new interesting possibilities which may
open up in the context of our model. The reason that we are
considering this speculation is that it may imply the possibility
of experimental investigation of the baryon number violation in these
experiments.  

  In the absence of a more complete treatment of the non-equilibrium
problem at hand, we will continue to use language of equilibrium
system to discuss even this possibility. Thus, from the resulting
energy densities, we will estimate an effective temperature $T_{eff}$.
This effective temperature, in the context of non-thermal symmetry 
restoration, will only represent whether the symmetry restoration can
happen, and if it does happen, how far the system is into the
symmetric phase, away from the transition point. Clearly all these
numbers are to be taken as only giving a crude picture of the
possibility of symmetry restoration. 
   
 In the absence of thermal equilibrium one needs some other estimate
for the size of the region. As we have mentioned, we will translate the
energy densities to an effective temperature $T_{eff}$ (which should
be  taken as giving a scale representing the properties of the effective
potential). It may not be an unreasonable guess then to take the
system size also to be given by the same scale, i.e. $T_{eff}^{-1}$.
Since our motivation is only to see if the densities are large enough
at least to be consistent with the restoration of electroweak symmetry,
we will take the size of the region to be given simply by
$T_{ew}^{-1}$, where $T_{ew} \simeq 100$ GeV is the transition
temperature for the electroweak symmetry breaking. With this, we will
again estimate the temperature $T_f$ (which we now denote as $T_{eff}$ 
since it should be taken as a parameter characterizing the
modification  of the effective potential) of the dense spot using 
Eq.(13), but now with $r_{eq} = T_{ew}^{-1}$.

 However, as we will see, in the case of LHC energies one gets
$T_{eff} > T_{ew}$ which we interpret as the possibility of
electroweak symmetry restoration. For such a situation, we also
include electroweak degrees of freedom in Eq.(13), i.e. we take $g_* =
100$. Also, we use the QCD values for the values of interface tension
$\sigma$ and false vacuum energy density $\rho$ in Eq.(13) when $T_{eff}$ 
comes out to be less than 100 GeV, and use the values (again, only in
Eq.(13)) typical of the electroweak scale, i.e., $\sigma \simeq T_{ew}^3$, and
$\rho \simeq T_{ew}^4$, when the $T_{eff}$ exceeds 100 GeV. There is a 
technical problem in solving for $T_{eff}$ in this manner from Eq.(13). When 
$T_{eff}$ is exactly equal to 100 GeV, then one must allow for 
the changeover in the value of $\rho$ from the value relevant to QCD to the 
value relevant for the electroweak symmetric vacuum (i.e., the latent heat 
of the electroweak phase transition). We take care of this by using the 
following prescription. Below, we will be giving plots of $T_{eff}$ as a 
function of $B^{1/4}$. As the value of $B^{1/4}$ is decreased, it 
corresponds to increasing value of $E_f$, and consecutively, 
increasing value of $T_{eff}$ (via Eq.(13)). 
For some cases, as $B^{1/4}$ is decreased, $T_{eff}$ increases from
a value below 100 GeV, and we reach a point when 
$T_{eff}$ just equals 100 GeV (for the size of the region equal to 
(100 GeV)$^{-1}$). For these values of $B^{1/4}$,  we use QCD values for 
$\rho^\prime$ and $\sigma^\prime$. As $B^{1/4}$ is decreased further, $E_f$
increases, but this increase is not enough to convert entire region of size
(100 GeV)$^{-1}$ into the electroweak symmetric phase (i.e., appropriate for
using electroweak values of $\rho^\prime$ and $\sigma^\prime$). However, a
region of somewhat smaller size can always be in the electroweak symmetric
phase. Equivalently, the region of size (100 GeV)$^{-1}$ will be in the mixed
phase. The temperature $T_{eff}$ will remain fixed equal to 100 GeV for these
values of $B^{1/4}$. Eventually a value of $B^{1/4}$ will be reached which
leads to large enough $E_f$ which can convert the whole region of size
(100 GeV)$^{-1}$ into the electroweak symmetric phase. Thus, in the plots
below, when $T_{eff}$ crosses the value 100 GeV, 
there will be a very small range of $B^{1/4}$ for which $T_{eff}$ 
will remain constant, equal to 100 GeV.

 As we discuss this possibility motivated by the non-thermal
 restoration of the electroweak symmetry, we will not present results
for the case of RHIC as there we find $T_{eff}$ to be always much less than
100 GeV. Since $T_{eff}$ should not be  interpreted as the
temperature of the hot spot, it is not clear what implications
one can obtain from this value of $T_{eff}$ for RHIC. 

 Fig.8 gives plots of $T_{eff}$ for different parameters for the case
$\Delta z_i = 1$. We see that for $A = 200$, $T_{eff}$ is well above 
$T_{ew} \simeq 100$ GeV (with ${\Delta z}_i$ = 1 fm) even for 
$\epsilon_i \simeq$ 1.3 TeV/fm$^3$ which is roughly the expected energy
density at the initial stage in Pb-Pb collision at $\sqrt{s} = 5.5$ TeV
at LHC. Such a large value of $T_{eff}$ indicates extremely dense
particle system, which may be taken as indicating a strong possibility
of non-thermal restoration of electroweak symmetry. 
Interesting thing is that even for rather small values of $A = 50$,
it is possible to get $T_{eff} > T_{ew}$ by using large values of $\sqrt{s}$.
This is important since early freezeout is more natural for small values
of $A$. Fig.9 shows similar plots for the case when $\Delta z_i$ in Eq.(4) is
equal to 0.15 fm. Resulting values of $T_{eff}$ are much smaller now, and for
$\sqrt{s}$ = 5.5 TeV resulting $T_{eff}$ is smaller than $T_{ew} \sim 100$ GeV.

\section{Discussion and Conclusions}

 The possibility of producing a false vacuum bubble in laboratory 
in the context of relativistic quantum field theory will have
important implications. Bubble wall propagation, its collapse and
eventual decay into particles are issues which have relevance for 
various phase transitions in the early universe. As we have argued,
such a false vacuum bubble should lead to a hot spot with temperature
which may be well in excess of the initial temperature of the QGP
system (as given by the initial energy density $\epsilon_i$). For
example, even for a reasonably conservative set of parameters, the
temperature of the hot spot can reach as high as 3 GeV for LHC (as
shown by the dashed plot in Fig.6 for $\sqrt{s}$ = 5.5 TeV case). 
One may get much higher
temperatures for more optimistic values of various parameters. Such a
hot spot will have many obvious signals. For example, one will 
expect increased production of heavy quarks. There should be 
anomalous production of very large $P_T$ partons, dileptons, and photons 
(depending on what is the distribution of the decay product from the decay of
colliding bubble walls). Of course if the bubble collapse
does not proceed to sizes much smaller than 1 fm (for example, due to 
quantum fluctuations on the bubble wall background, or if the
center of the shell is not entirely devoid of partons, or if the bubble
breaks up due to asphericity before collapsing to very small sizes), 
the resulting temperature for this 1 fm size hot spot may not be large
enough to lead to clean signals.

  We have also speculated on the possibility that due to dense parton
system resulting from the decay of bubble walls, one may get
non-thermal restoration of electroweak symmetry. If that happens then
it may open up the possibility of observing unsuppressed baryon number 
violation via sphaleron processes \cite{sphl0,sphl1,sphl2}.  An important 
issue in this regard is the size of the region as compared to the
sphaleron size. Below $T_{ew}$, when 
electroweak symmetry is broken, sphaleron is a solution of classical 
equations of motion. Baryon number violating processes are dominated by 
sphalerons \cite{sphl1} with a size of order (3 GeV)$^{-1}$. This is a very 
large region compared to the size we have considered above, i.e. 
(100 GeV)$^{-1}$. If we consider the size of the hot spot to be about 
(3 GeV)$^{-1}$, then resulting values of $T_{eff}$  never 
exceed about 15 GeV.
It is not clear what really should be the lower limit for the size of the 
region for sphaleron interactions to occur in the 
symmetric phase \cite{sphl1,sphl2}. It is possible that in the symmetric 
phase, sphaleron processes in smaller regions may not be too suppressed
\cite{sphl1}. Even in the symmetry broken phase, the core of sphaleron is 
only about (20 GeV)$^{-1}$ large \cite{sphl2}. For hot spot of size $\sim$ 
(20 GeV)$^{-1}$, we find that $T_{eff}$ is less than about 65 GeV, even with
$ A = 200$, and $\sqrt{s}$ = 30 TeV (with ${\Delta z}_i$ = 1 fm).  
 
 In conclusion, we have discussed the possibility that under certain
situations, when a shell like expanding parton system emerges from
the collision of ultra-relativistic nuclei, a bubble of pure false vacuum
may be left behind. We find that the net energy of this bubble may 
be a significant fraction of the total energy of the initial parton system. 
This bubble undergoes free, relativistic collapse. Due to extremely large 
Lorentz contraction factor, the bubble wall thickness decreases faster than 
the bubble radius. Due to this, bubble contraction proceeds down to very small
scales, much smaller than the typical QCD scale of 1 fm (assuming
that any quantum fluctuations on the background of the collapsing
ultra-relativistic wall do not dissipate most of its energy when bubble
radius is about 1 fm). Eventually different portions of bubble
wall collide, converting all of their kinetic energy (which equals the
initial bubble energy) into particles. These particles may eventually
thermalize, leading to a hot spot. We have estimated the expected temperatures
in this hot spot and find that it can be well above the initial temperature
of the QGP system. There will be clear signals of
such hot spots, such as increased production of heavy quarks, very large
$P_T$ partons, dileptons, photons etc.

\vskip .2in
\centerline {\bf ACKNOWLEDGEMENTS}
\vskip .1in

  We are very thankful to Pankaj Agrawal, Sanatan Digal, Avijit Ganguly,
Amit Kundu, Biswanath Layek, Shashi Phatak, and Supratim Sengupta for 
useful discussions and comments.

%%%%%%%%%%%%%%%%%%% 

\newpage

%%%%%%%%%%%%%%%%%%%%%%%%%%%%%%%%%%%%%%%%%%%%%%%%%%%%%%%%%%%%%%%%
%\vskip -0.75in
\begin{figure}[h]
\begin{center}
\leavevmode
\epsfysize=15truecm \vbox{\epsfbox{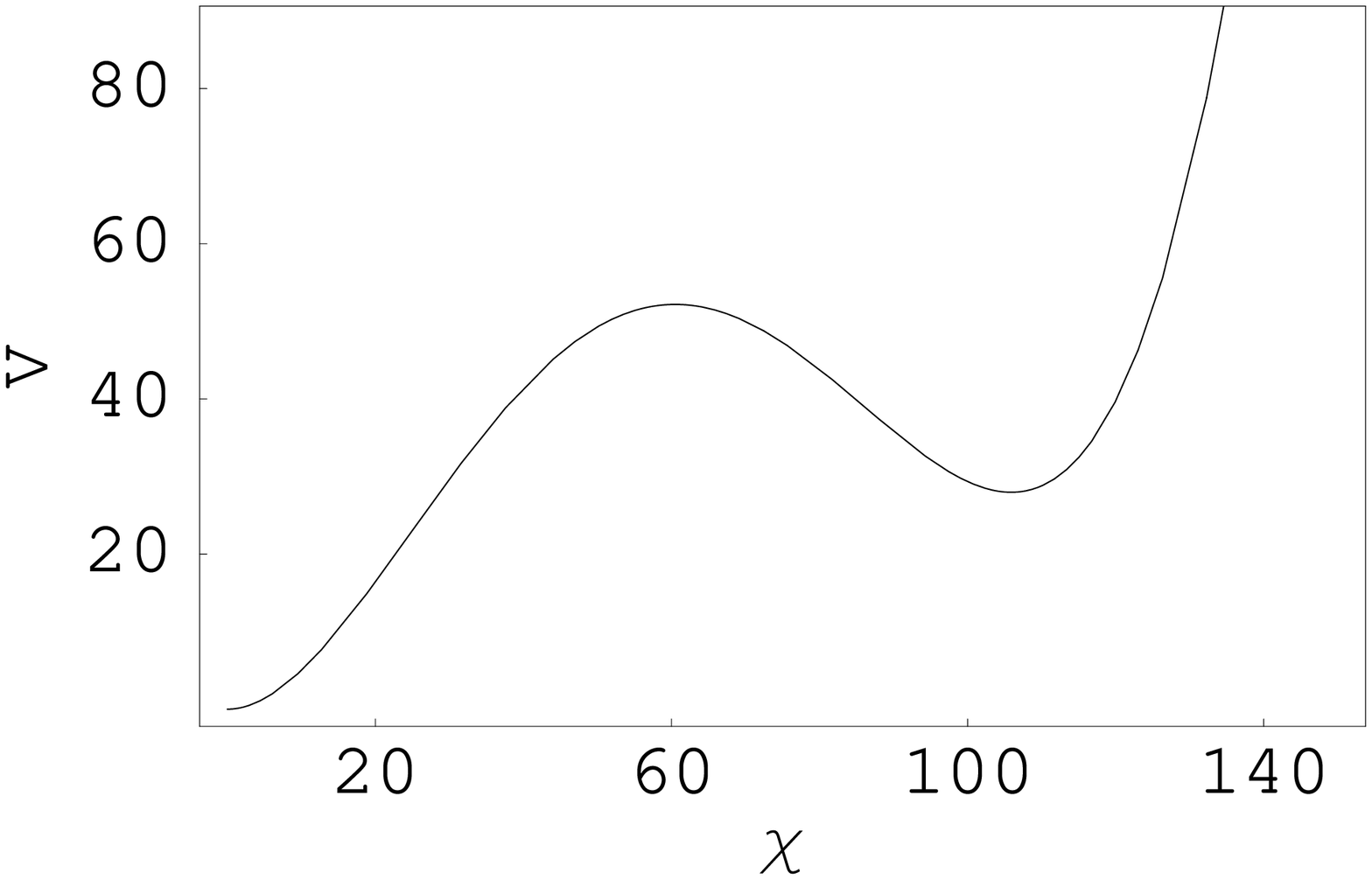}}
\end{center}
\vskip -1.5in
\caption{}{Plot of $V(\chi)$ (in MeV/fm$^3$) vs. $\chi$ (in MeV), as 
given in Eq.(3). The absolute minimum at $\chi = 0$ corresponds to the 
confining true vacuum while the local minimum at $\chi \ne 0$ corresponds 
to the metastable deconfining vacuum.}
\label{Fig.1}
\end{figure}
%%%%%%%%%%%%%%%%%%%%%%%%%%%%%%%%%%%%%%%%%%%%%%%%%%%%%%%%%%%%%%%%%%

%%%%%%%%%%%%%%%%%%%%%%%%%%%%%%%%%%%%%%%%%%%%%%%%%%%%%%%%%%%%%%%%
%\vskip -0.25in
\begin{figure}[h]
\begin{center}
\leavevmode
\epsfysize=12truecm \vbox{\epsfbox{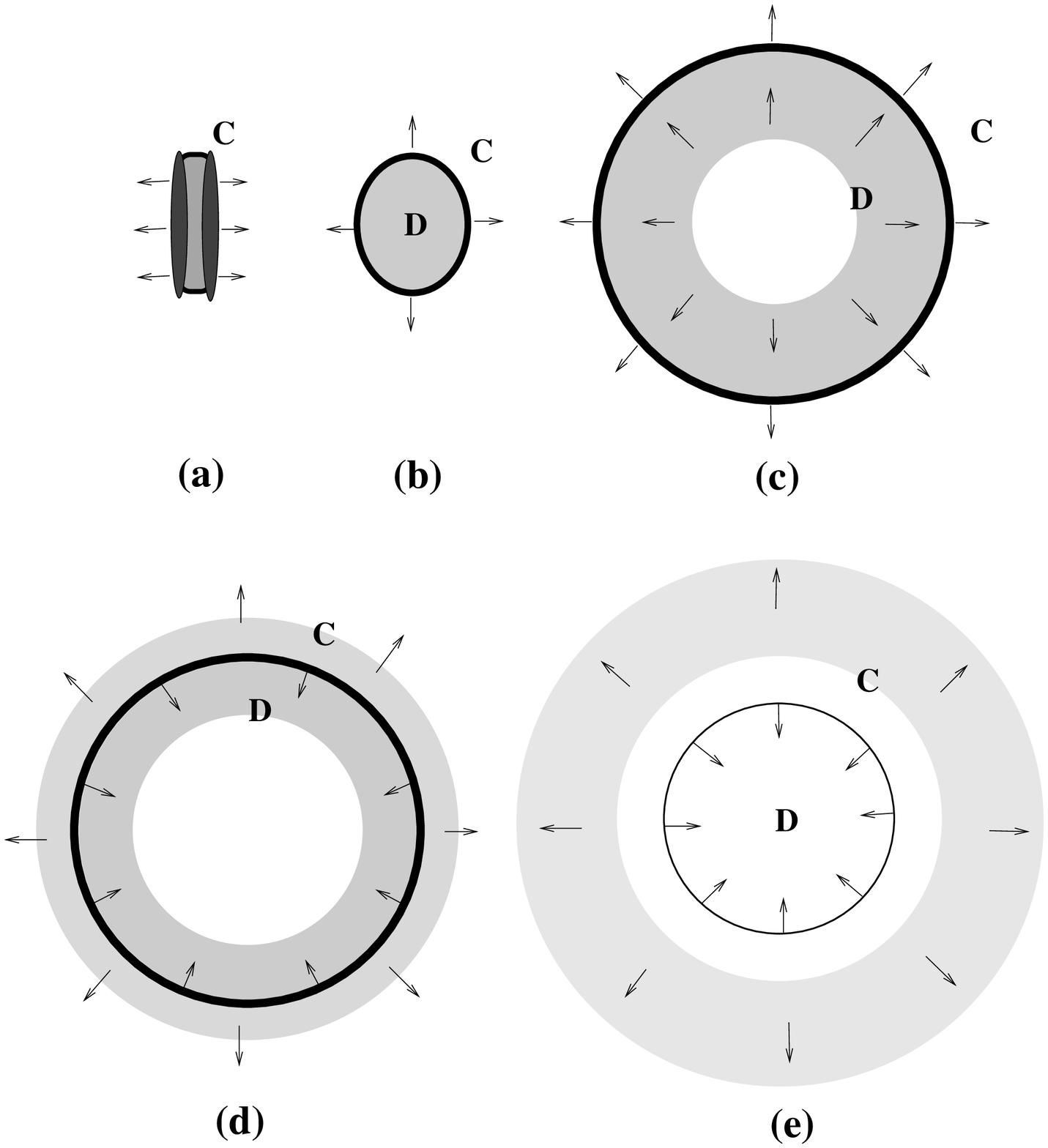}}
\end{center}
%\vskip -0.3cm
\caption{}{(a) The initial stage showing beginning of longitudinal
expansion of the parton system between the two nuclei receding after 
overlap. (b) Beginning of transverse expansion. C and D denote regions 
with confining vacuum and deconfining vacuum, respectively. Thick solid
line at the boundary denotes the interface separating the two vacua.
(c) Development of shell structure with depletion of partons in the 
center due to expansion. (d) Stage of hadronization of the 
parton shell as the interface shrinks 
through the shell. (e) Relativistic collapse of the interface. Interface 
is shown to be thinner due to Lorentz contraction.}
\label{Fig.2}
\end{figure}
%%%%%%%%%%%%%%%%%%%%%%%%%%%%%%%%%%%%%%%%%%%%%%%%%%%%%%%%%%%%%%%%%%
%%%%%%%%%%%%%%%%%%%%%%%%%%%%%%%%%%%%%%%%%%%%%%%%%%%%%%%%%%%%%%%%
%\vskip -0.3in
\begin{figure}[h]
\begin{center}
\leavevmode
\epsfysize=8truecm \vbox{\epsfbox{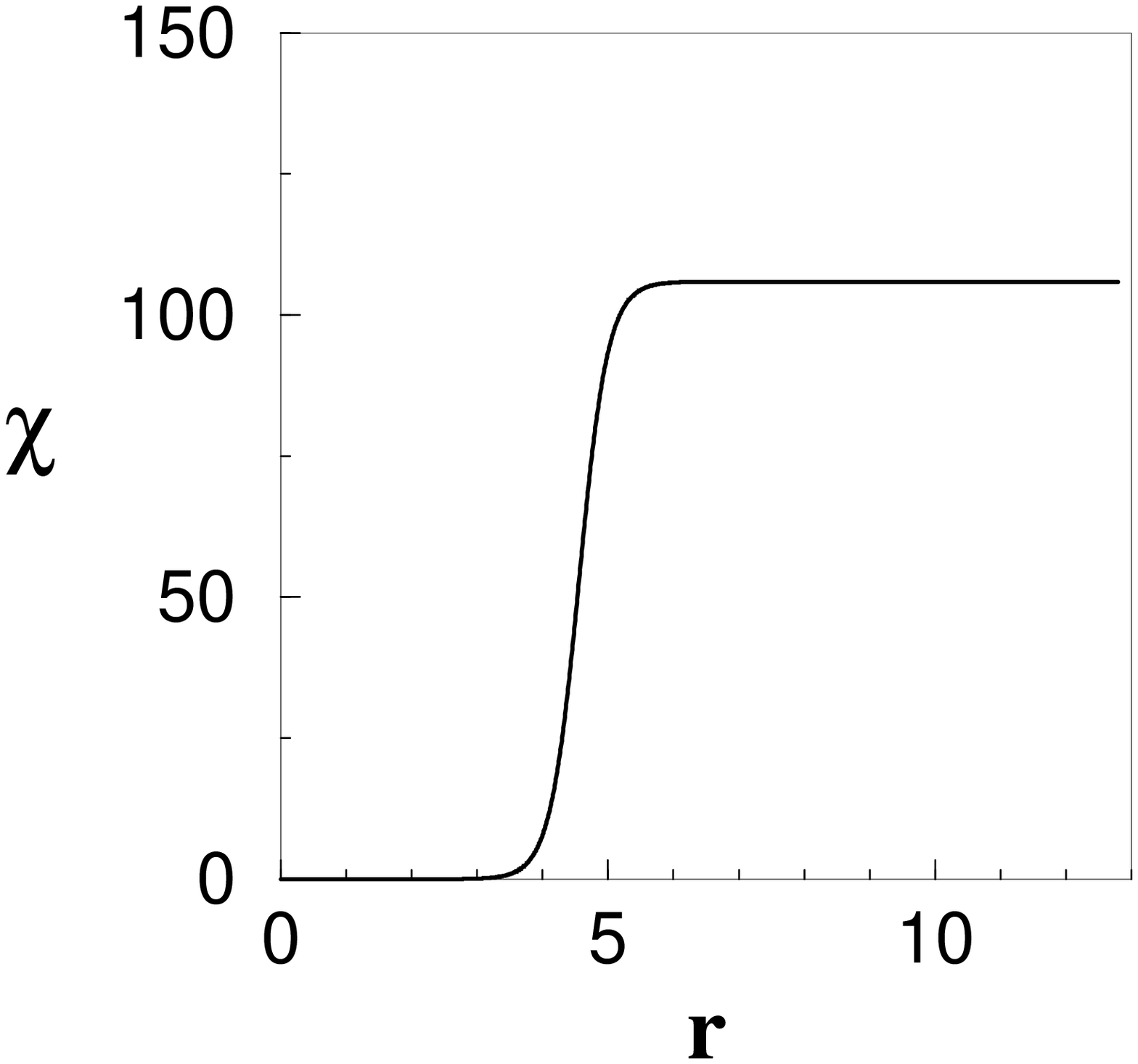}}
\epsfysize=8truecm \vbox{\epsfbox{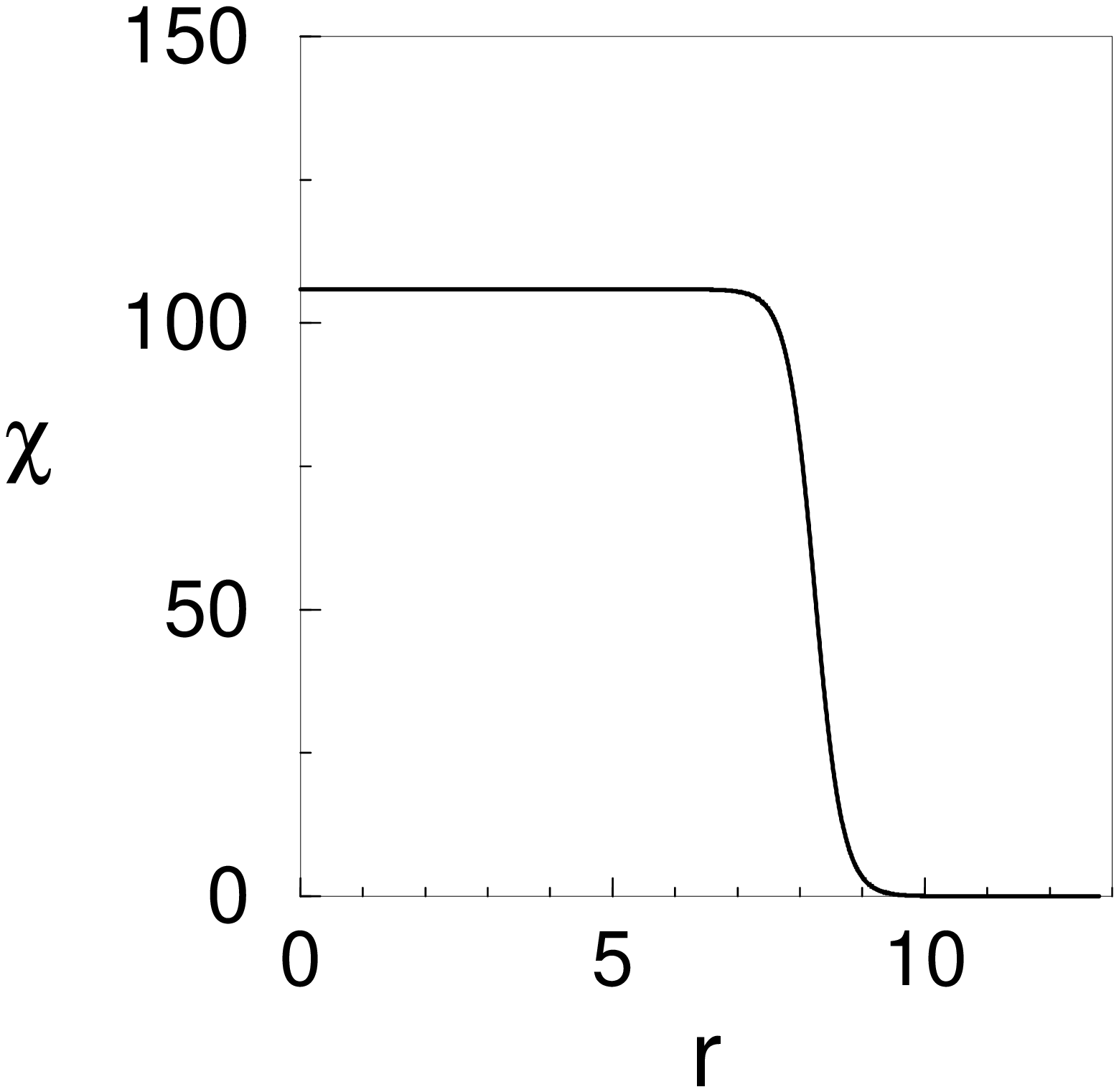}}
\end{center}
%\vskip -0.3cm
\caption{}{Left figure shows the true vacuum quantum bubble obtained
as the solution of Eq.(5). Right figure shows the false vacuum bubble
obtained by inverting the profile of the true vacuum bubble. $\chi$ is 
in MeV and $r$ is in fm.}
\label{Fig.3}
\end{figure}
%%%%%%%%%%%%%%%%%%%%%%%%%%%%%%%%%%%%%%%%%%%%%%%%%%%%%%%%%%%%%%%%%%
\newpage
%%%%%%%%%%%%%%%%%%%%%%%%%%%%%%%%%%%%%%%%%%%%%%%%%%%%%%%%%%%%%%%%
%\vskip -1.00in
\begin{figure}[h]
\begin{center}
\leavevmode
\epsfysize=10truecm \vbox{\epsfbox{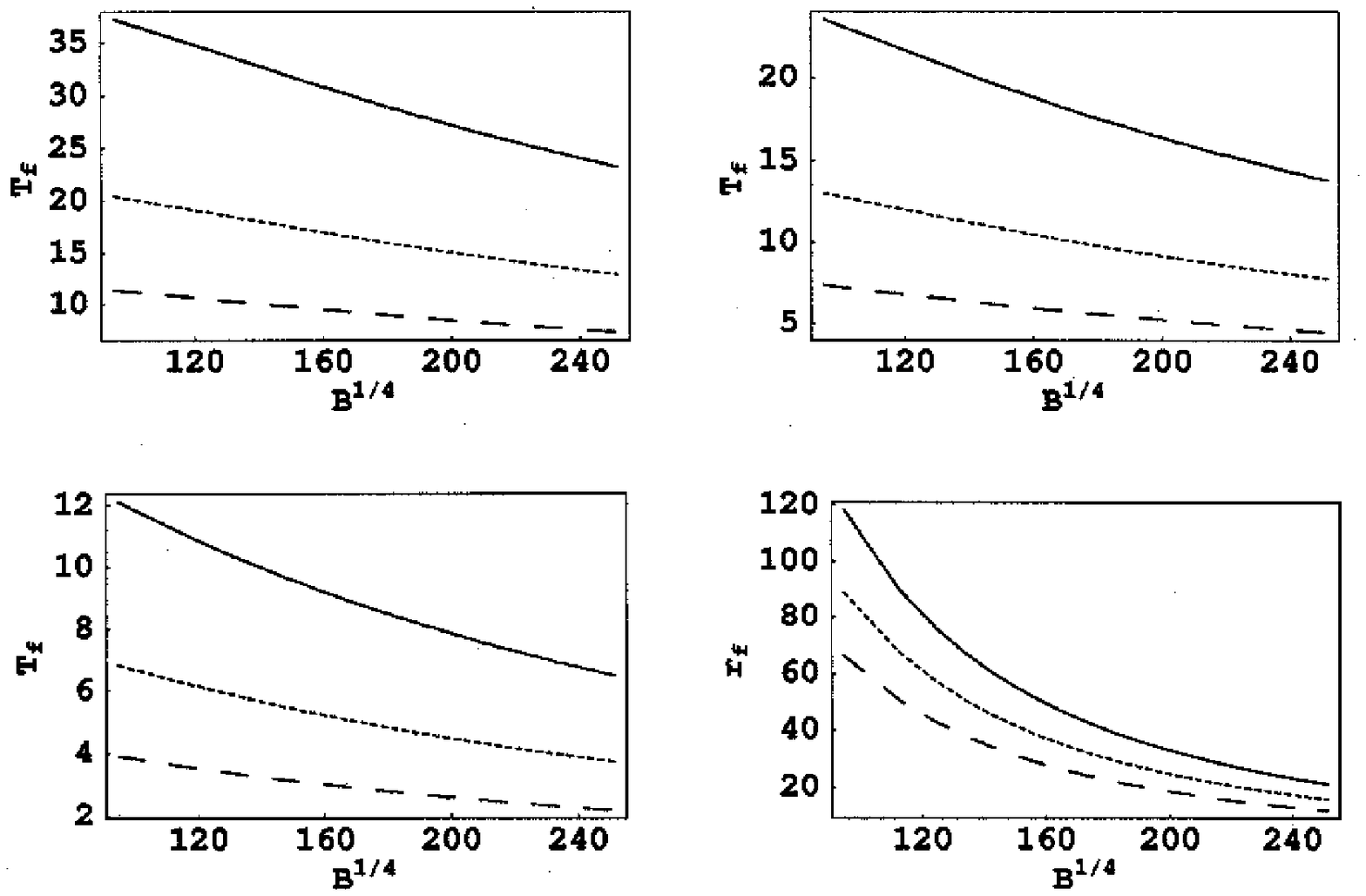}}
\end{center}
%\vskip -4.00in
\caption{}{Figures on top left and top right correspond
to $\sqrt{s}$ = 30 and 15 TeV, respectively. Bottom two figures
correspond to $\sqrt{s}$ = 5.5 TeV. $T_f$ is in GeV, $B^{1/4}$ in MeV,
and $r_f$ is in fm. Solid, dotted, and dashed curves correspond to
$A$ = 200, 100, and 50 respectively. These plots correspond to
$\Delta z_i = 1$ fm in Eq.(4). These plots are for $r_{eq}$
given in Eq.(10) corresponding to the case when only gluons thermalize.}
\label{Fig.4}
\end{figure}
%%%%%%%%%%%%%%%%%%%%%%%%%%%%%%%%%%%%%%%%%%%%%%%%%%%%%%%%%%%%%%%%%%

%%%%%%%%%%%%%%%%%%%%%%%%%%%%%%%%%%%%%%%%%%%%%%%%%%%%%%%%%%%%%%%%
%\vskip -1.00in
\begin{figure}[h]
\begin{center}
\leavevmode
\epsfysize=10truecm \vbox{\epsfbox{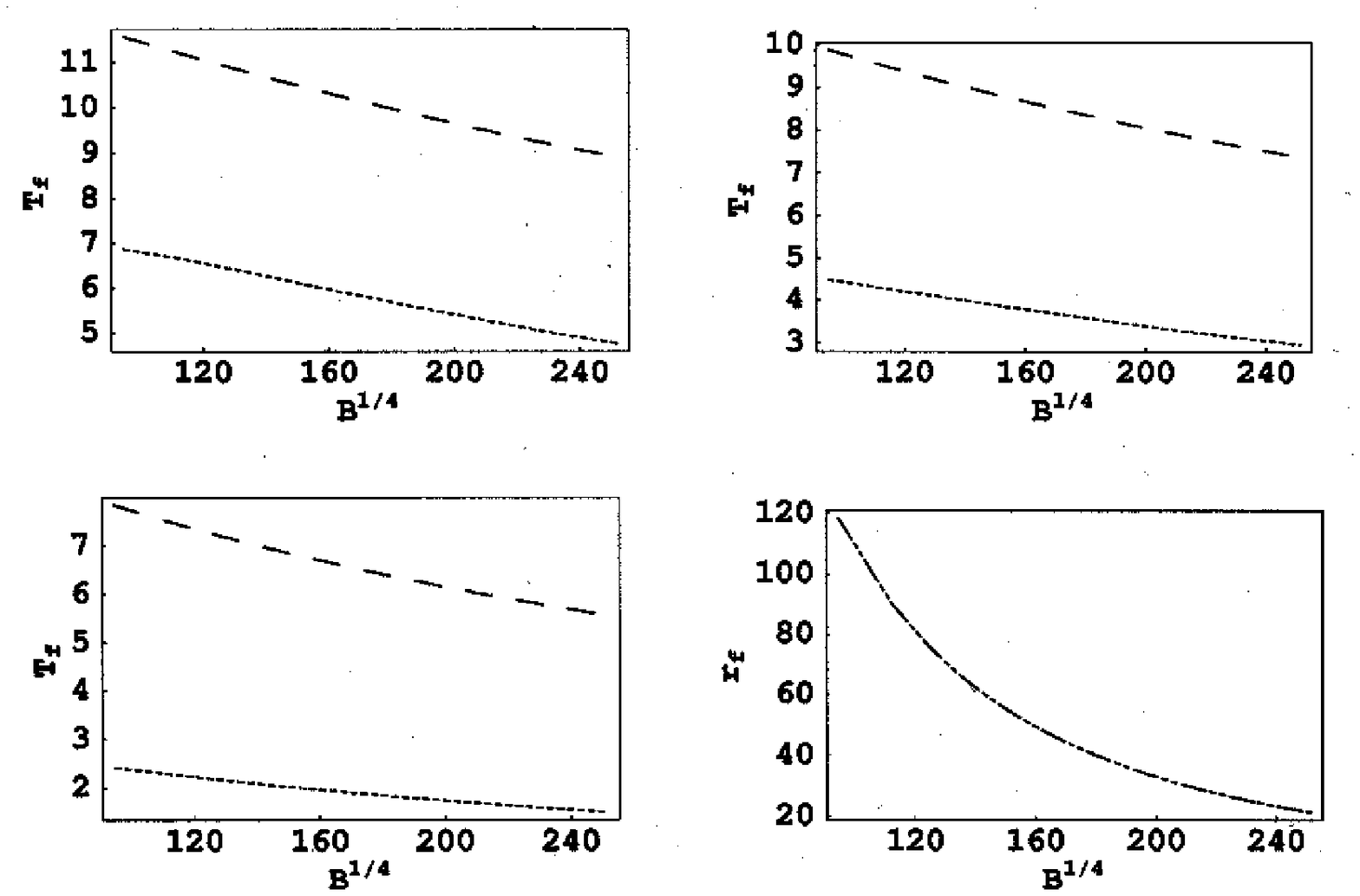}}
\end{center}
%\vskip -4.00in
\caption{}{All plots in this figure correspond to $A = 200$. 
Figures on top left and top right correspond
to $\sqrt{s}$ = 30 and 15 TeV, respectively. Bottom two figures
correspond to $\sqrt{s}$ = 5.5 TeV. $T_f$ is in GeV, $B^{1/4}$ in MeV,
and $r_f$ is in fm. Dotted and dashed curves correspond to the two
cases, $r_{eq} = \alpha_s T$ (Eq.(11)), and $r_{eq} = 0.15$ fm
respectively. Plot of $r_f$ shows bubble radius for both these cases.
(It is the same as the plot of $r_f$ shown by the solid curve
in Fig.4.)
These plots correspond to $\Delta z_i = 1$ fm in Eq.(4).}
\label{Fig.5}
\end{figure}
%%%%%%%%%%%%%%%%%%%%%%%%%%%%%%%%%%%%%%%%%%%%%%%%%%%%%%%%%%%%%%%%%

%%%%%%%%%%%%%%%%%%%%%%%%%%%%%%%%%%%%%%%%%%%%%%%%%%%%%%%%%%%%%%%%
\begin{figure}[h]
\begin{center}
\leavevmode
\epsfysize=10truecm \vbox{\epsfbox{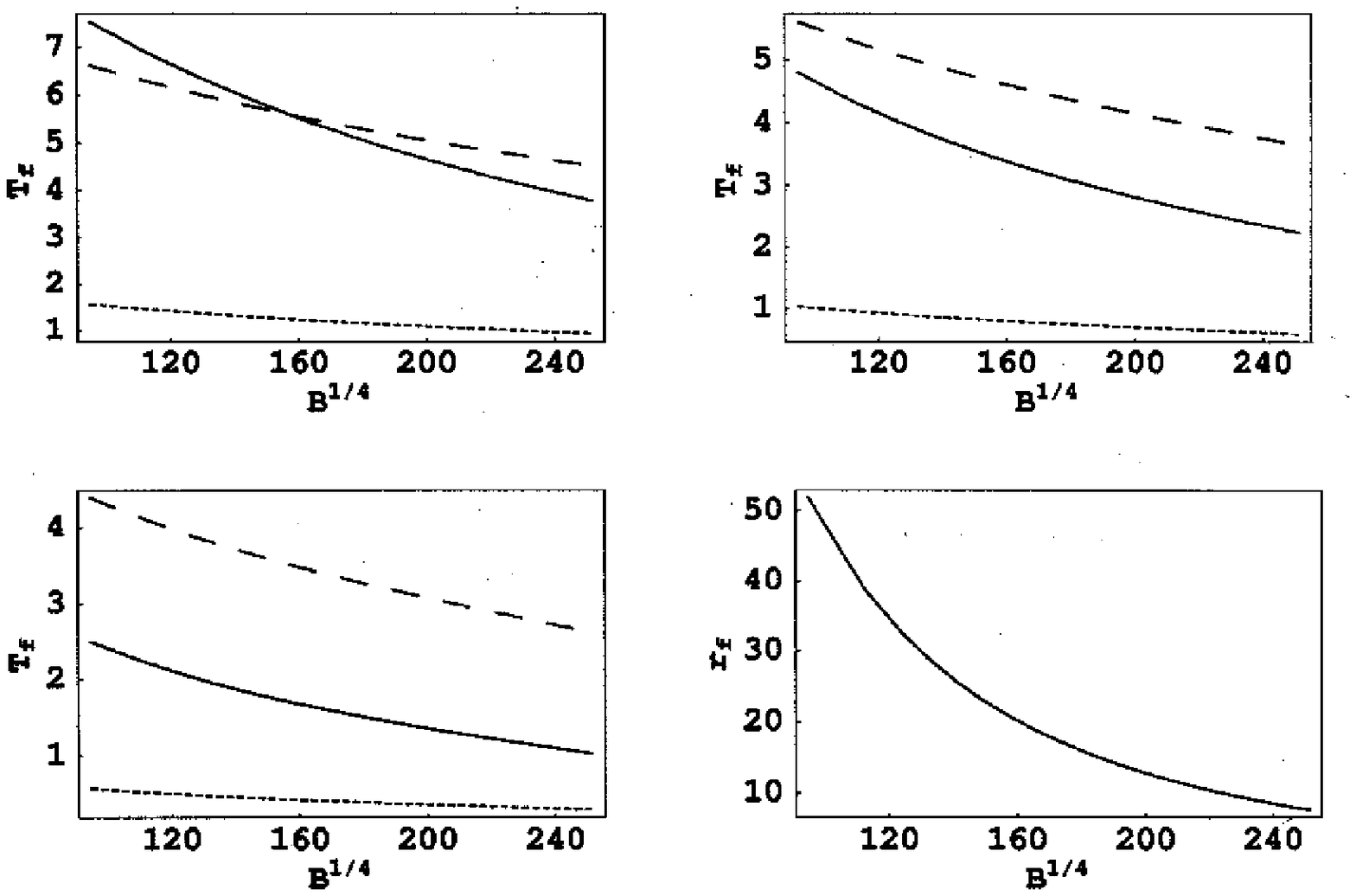}}
\end{center}
%\vskip -4.00in
\caption{}{Similar plots as in Fig.5, but now with $\Delta z_i = 0.15$ 
fm. Here we show three plots in each figure for $T_f$. Solid, dotted, 
and dashed plots correspond to the three choices for $r_{eq}$ as given by 
Eq.(10)-(12), respectively.} 
\label{Fig.6}
\end{figure}
%%%%%%%%%%%%%%%%%%%%%%%%%%%%%%%%%%%%%%%%%%%%%%%%%%%%%%%%%%%%%%%%%

%%%%%%%%%%%%%%%%%%%%%%%%%%%%%%%%%%%%%%%%%%%%%%%%%%%%%%%%%%%%%%%%
%\vskip -1.00in
\begin{figure}[h]
\begin{center}
\leavevmode
\epsfysize=10truecm \vbox{\epsfbox{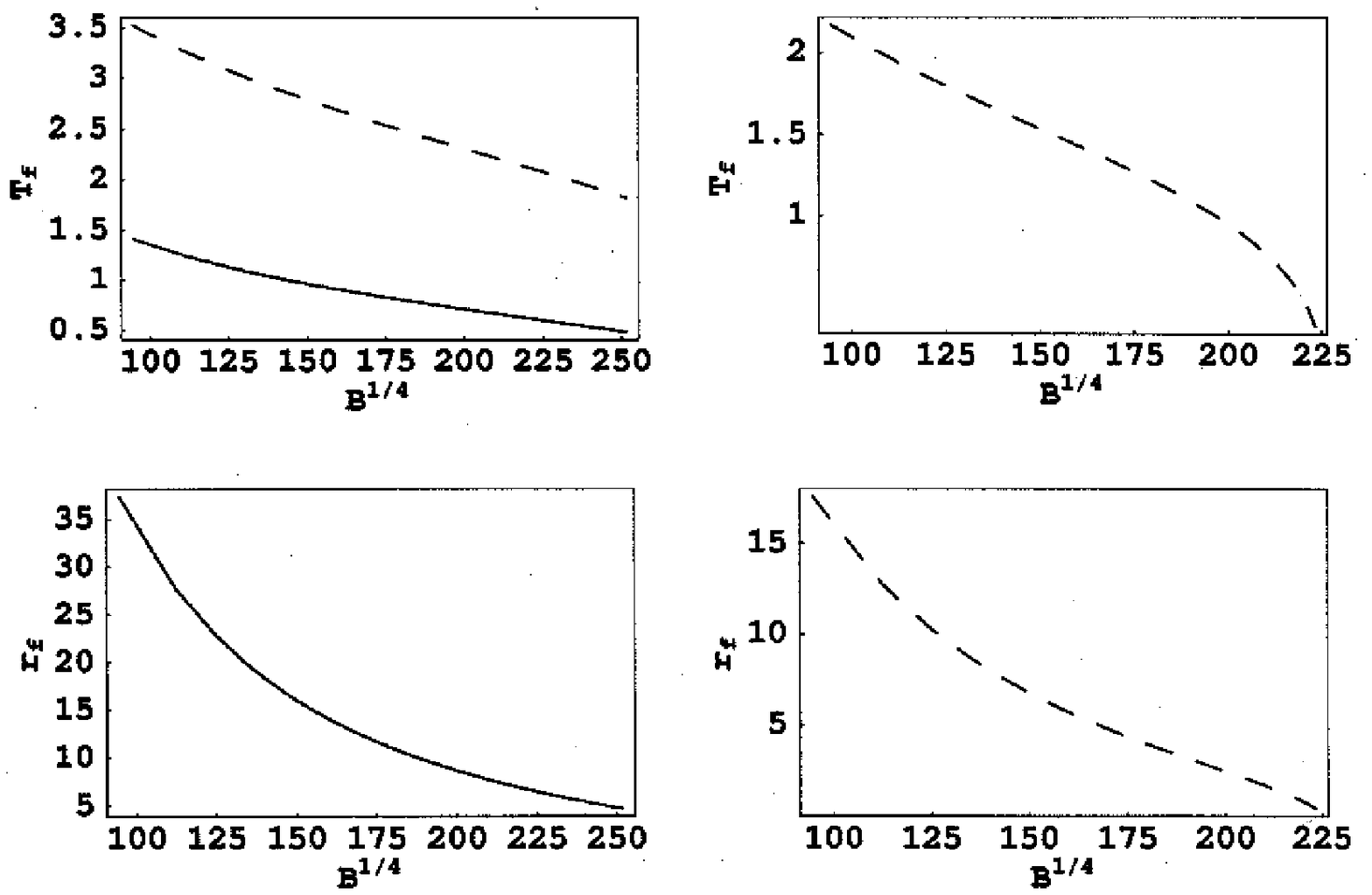}}
%\vskip -4.00in
\end{center}
\caption{}{Plots for $\sqrt{s}$ = 200 GeV. Figures on top left and 
top right correspond to $\Delta z_i$ (in Eq.(4)) = 1 fm and 0.22 fm
respectively. Figures below these give corresponding plots of the shell
radius $r_f$. For $\Delta z_i = 1$ fm case, solid and dashed plots 
correspond to $r_{eq}$ given by Eq.(10) and Eq.(12) respectively. No
solution is found for the case when $r_{eq} = \alpha_s T$ (Eq.(11))
for reasonable values of $B^{1/4}$. For $\Delta z_i = 0.22$ fm,
solutions for reasonable values of $B^{1/4}$ are 
found only for the case when $r_{eq} = 0.15$ fm,
as shown by the dashed plot.  $A = 200$ for all these plots.} 
\label{Fig.7}
\end{figure}
%%%%%%%%%%%%%%%%%%%%%%%%%%%%%%%%%%%%%%%%%%%%%%%%%%%%%%%%%%%%%%%%%%

%%%%%%%%%%%%%%%%%%%%%%%%%%%%%%%%%%%%%%%%%%%%%%%%%%%%%%%%%%%%%%%%
\vskip -1.00in
\begin{figure}[h]
\begin{center}
\leavevmode
\epsfysize=10truecm \vbox{\epsfbox{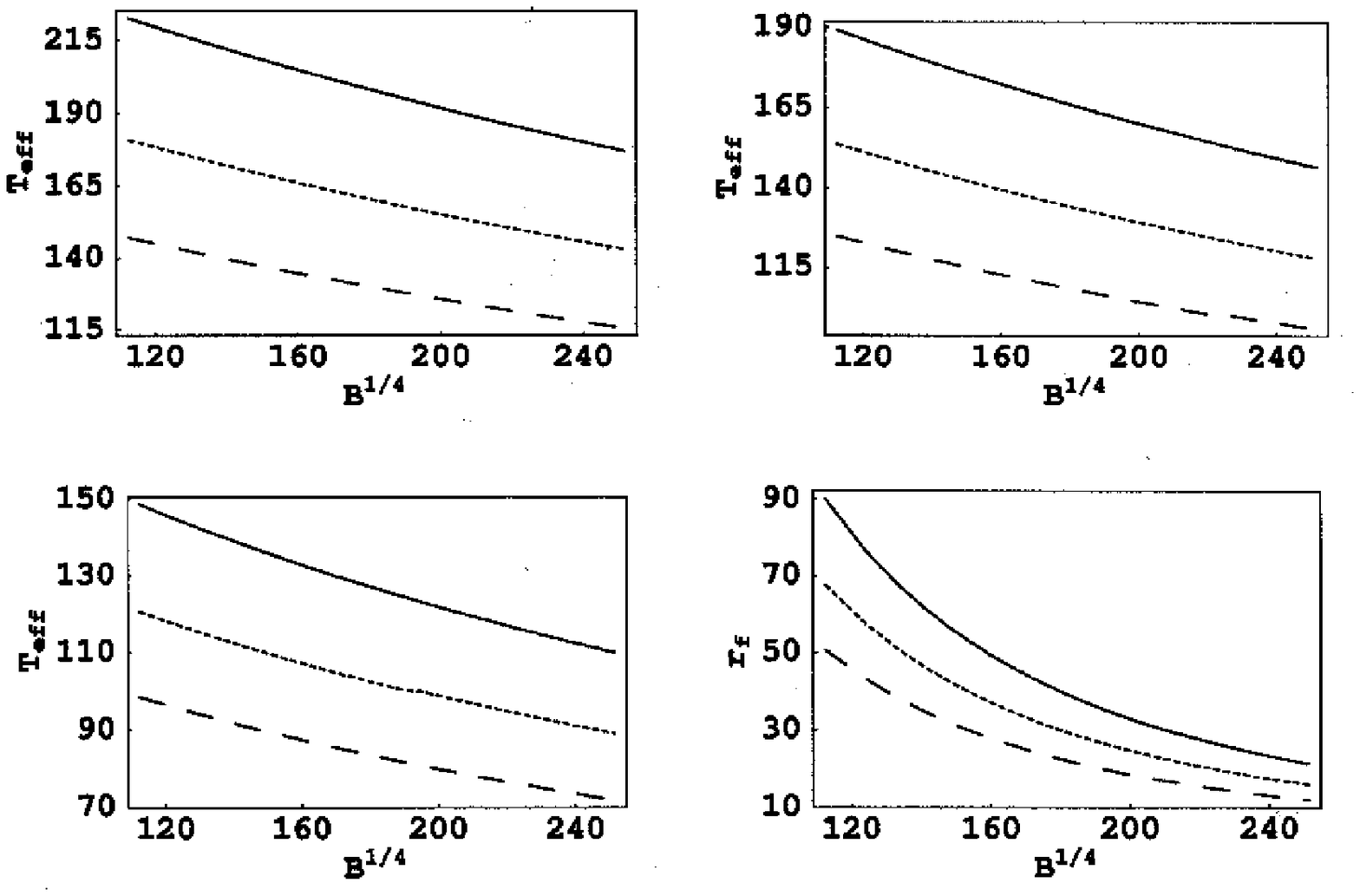}}
\end{center}
%\vskip -4.00in
\caption{}{Figures on top left and top right correspond
to $\sqrt{s}$ = 30 and 15 TeV, respectively. Bottom two figures
correspond to $\sqrt{s}$ = 5.5 TeV. $T_{eff}$ is in GeV, $B^{1/4}$ in MeV,
and $r_f$ is in fm. Solid, dotted, and dashed curves correspond to
$A$ = 200, 100, and 50 respectively. These plots correspond to
$\Delta z_i = 1$ fm in Eq.(4). (Note that the range of $B^{1/4}$
in the plots here and in Fig.9 is slightly different from the range 
in Figs.4-7.)}
\label{Fig.8}
\end{figure}
%%%%%%%%%%%%%%%%%%%%%%%%%%%%%%%%%%%%%%%%%%%%%%%%%%%%%%%%%%%%%%%%%%

%%%%%%%%%%%%%%%%%%%%%%%%%%%%%%%%%%%%%%%%%%%%%%%%%%%%%%%%%%%%%%%%
\vskip -1.00in
\begin{figure}[h]
\begin{center}
\leavevmode
\epsfysize=10truecm \vbox{\epsfbox{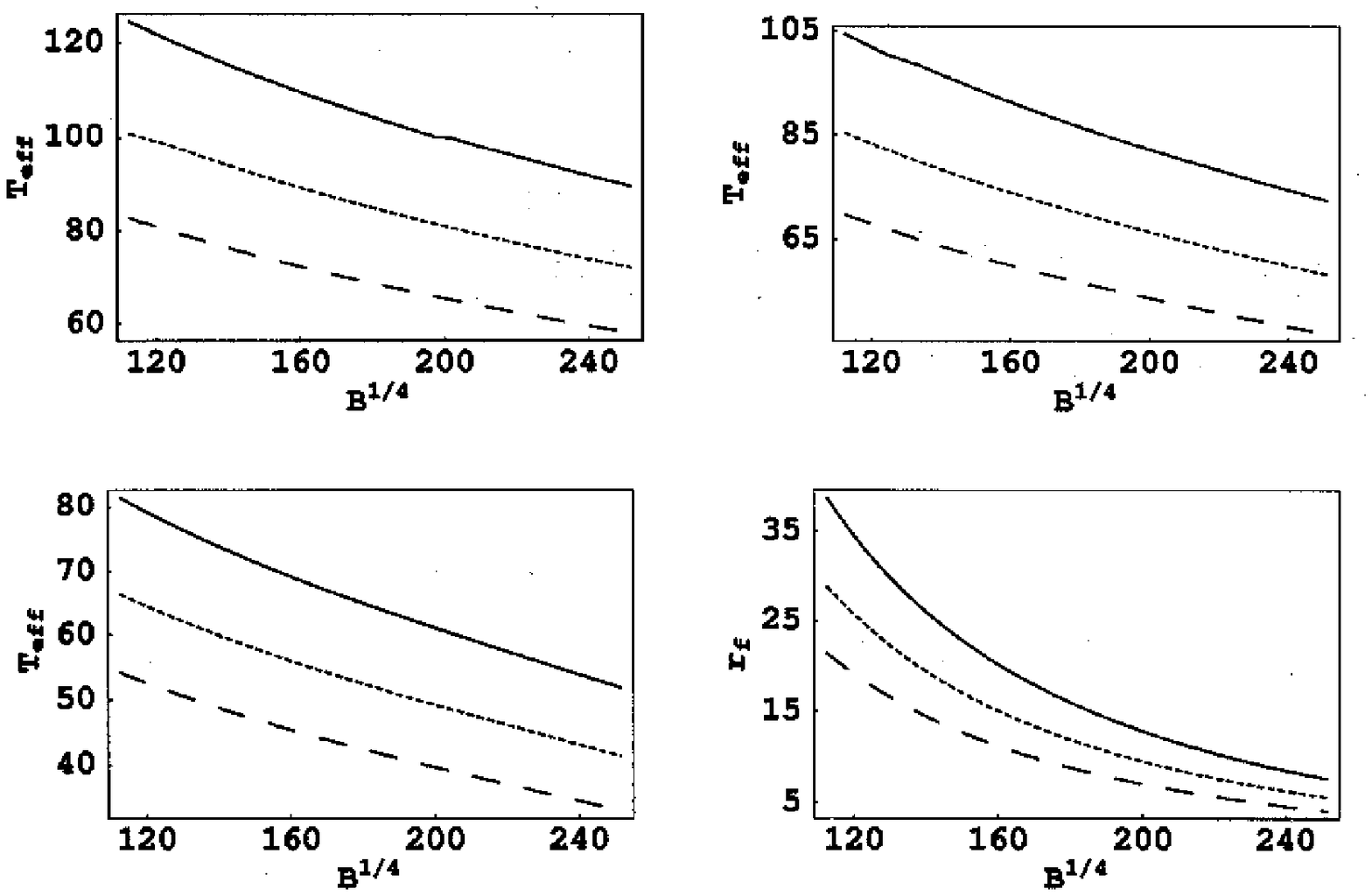}}
\end{center}
%\vskip -4.00in
\caption{}{Same plots as in Fig.8, but with
$\Delta z_i = 0.15$ fm in Eq.(4).}
\label{Fig.9}
\end{figure}
%%%%%%%%%%%%%%%%%%%%%%%%%%%%%%%%%%%%%%%%%%%%%%%%%%%%%%%%%%%%%%%%%%

%\end{multicols}
\end{document}